\documentclass[a4paper,11pt]{article}
\pdfoutput=1 

\usepackage{jheppub} 

\usepackage[T1]{fontenc} 
\usepackage{lipsum}
\usepackage{graphicx} 
\usepackage{multirow}
\usepackage{lmodern}
\usepackage{amsmath,amssymb}
\usepackage{mathtools}
\usepackage[numbers,sort&compress]{natbib}
\usepackage{pgfplots,pgfplotstable}
\usepgfplotslibrary{fillbetween}
\pgfplotsset{compat=newest}
\usetikzlibrary{calc}
\usepackage{lipsum}
\usepackage{stackrel}
\usepackage{cleveref}
\usepackage{ifthen}
\usepackage{xspace}
\usepackage[shortlabels]{enumitem}
\usepackage{cancel}
\usepackage{slashed}
\usepackage{arydshln}
\usepackage{url}
\usepackage{hyperref}
\usepackage{dsfont}
\newcommand{\MPQ}{M_{\cancel{\text{\tiny PQ}}}}

\title{\boldmath Final state rescattering effects in axio-hadronic $\eta$ and $\eta^{\prime}$ decays}

\author[a]{Daniele S. M. Alves}
\author[a,b,c]{Sergi Gonz\`{a}lez-Sol\'is}
\affiliation[a]{Theoretical Division, Los Alamos National Laboratory, Los Alamos, NM 87545, U.S.A.}
\affiliation[b]{Departament de F\'isica Qu\`{a}ntica i Astrof\'isica, Universitat de Barcelona,\\Mart\'i i Franqu\`{e}s, 1, 08028 Barcelona, Spain}
\affiliation[c]{Institut de Ci\`{e}ncies del Cosmos, Universitat de Barcelona,\\Mart\'i i Franqu\`{e}s, 1, 08028 Barcelona, Spain}

\preprint{LA-UR-24-20793}

\emailAdd{spier@lanl.gov}
\emailAdd{sergig@icc.ub.edu}

\abstract{It has been long-understood that final state rescattering effects provide $\mathcal{O}(1)$ corrections to hadronic meson decays rates, such as $\eta\to\pi\pi\pi$ and $\eta^{\prime}\to\eta\pi\pi$. Hence, one would expect that such effects would be just as important in axio-hadronic $\eta$ and $\eta^{\prime}$ decays, such as $\eta^{(\prime)}\to\pi\pi a$, where $a$ is an axion or axion-like particle (ALP). And indeed they are, as we show in this paper by using the treatment of dispersion relations to include the effects of strong final state interactions in several axio-hadronic processes, namely, $\eta^{(\prime)}\to\pi^0\pi^0 a$, $\eta^{(\prime)}\to\pi^+\pi^- a$, and $\eta^{\prime}\to\eta\pi^0 a$. We also compute the perturbative, leading order decay rates for multiple ALP emission, such as in $\eta^{(\prime)}\to\pi^{0}aa$, $\eta^{\prime}\to\eta aa$ and $\eta^{(\prime)}\to aaa$, and briefly discuss the expected corrections from strong interactions and the processes that must be considered for an accurate rate estimation of these multi-ALP decay channels.
}

\begin{document} 
\maketitle
\flushbottom

\section{Introduction}

Axions and axion-like particles (ALPs) are natural and generic predictions of theories beyond the Standard Model (BSM) with spontaneously broken PQ symmetries, and possess incredibly rich phenomenology \cite{Hagmann:2008zz,Workman:2022ynf}. In particular, axions/ALPs can couple to quarks, leptons, and gauge bosons, and their masses and decay constants can range over orders of magnitude, from $10^{-22}$\;eV to the Planck scale \cite{Kim:1986ax,Cheng:1987gp,Kim:2008hd,Marsh:2015xka}. In this vast parameter space, this class of particles can leave imprints in cosmology \cite{Cadamuro:2011fd,Millea:2015qra,Hoof:2017ibo,Depta:2020wmr}, modify properties of SM particles through virtual corrections \cite{Izaguirre:2016dfi,Choi:2017gpf,Alves:2017avw,Merlo:2019anv,Bauer:2019gfk,Cornella:2019uxs,MartinCamalich:2020dfe,Bauer:2021mvw,Bandyopadhyay:2021wbb}, mediate interactions between matter in the SM and/or in a dark sector \cite{Hostert:2020xku,Agrawal:2021dbo,Antel:2023hkf}, and be produced in reactions and decays in intensity frontier experiments \cite{Armengaud:2014gea,Alekhin:2015byh,Dolan:2017osp,FASER:2018eoc,Curtin:2018mvb,Dobrich:2019dxc,Beacham:2019nyx,Dusaev:2020gxi,Kelly:2020dda,Bai:2021gbm,Apyan:2022tsd}, high energy colliders \cite{Jaeckel:2012yz,Mimasu:2014nea,Jaeckel:2015jla,Knapen:2016moh,Brivio:2017ije,Bauer:2017ris,Bauer:2018uxu,Gavela:2019cmq}, and astrophysical environments \cite{Raffelt:2006cw,Payez:2014xsa,Jaeckel:2017tud,Chang:2018rso}.

Much attention has been given in the recent literature to ALPs that couple primarily to electroweak gauge bosons and/or leptons. The phenomenology of such ``electroweak ALPs''  and ``leptonic ALPs'' is amenable to perturbative calculations, such that signal predictions are generally accurate and reliable. On the other hand, low energy signals of hadronically-coupled axions and ALPs---and this includes the QCD axion that solves the strong CP problem \cite{Peccei:1977hh,Peccei:1977ur,Weinberg:1977ma,Wilczek:1977pj}---have been explored to a lesser degree \cite{Alves:2017avw,Aloni:2018vki,Bauer:2020jbp,Gan:2020aco,Alves:2020xhf,Bauer:2021wjo,Ertas:2020xcc,Blinov:2021say,Jerhot:2022chi,DiLuzio:2022tbb,Gao:2022xqz,Cornella:2023kjq}, in part due to the significant challenges in making accurate predictions. Specifically, the low energy phenomenology of hadronic axions/ALPs is plagued by large uncertainties associated with non-perturbative strong dynamics. While in the 1980s there were some attempts to fold these uncertainties into phenomenological predictions as ``fudge'' factors \cite{Antoniadis:1981zw,Bardeen:1986yb}, this has been largely forgotten in the recent literature, and, to a good approximation, studies of hadronic axion/ALP phenomenology in the past two decades have completely ignored any effects beyond naive leading order in chiral perturbation theory ($\chi$PT).

In this work, we attempt to mitigate this issue in the context of axio-hadronic decays of the $\eta$ and $\eta^{\prime}$ mesons. Unlike other mesons such as Kaons, $D$, and $B$ mesons---which decay through flavor-violating electroweak interactions---the decay of $\eta$ and $\eta^{\prime}$ mesons is dominated by the strong interactions. Furthermore, any axion or ALP that couples in the ultraviolet (UV) to quarks and/or gluons {\it must} be emitted in $\eta$ and $\eta^{\prime}$ decays. This makes these mesons an ideal choice as a first place to investigate the effects of low energy strong dynamics in axion/ALP phenomenology. In addition, the prospect of upcoming $\eta/\eta^{\prime}$ factories\footnote{The Jefferson Lab Eta Factory will improve existing measurements of rare $\eta/\eta^{\prime}$ decays by a few orders of magnitude, and the proposed REDTOP experiment aims to produce an unprecedented sample of $10^{14}(10^{12})$ $\eta(\eta^{\prime})$ mesons. These will offer unique opportunities to probe BSM physics in the unflavored meson sector.}, such as the Jefferson Lab Eta Factory (JEF)~\cite{JEF} and REDTOP~\cite{REDTOP:2022slw}, further motivates the investigation of rare $\eta^{(\prime)}$ decays as probes of BSM physics.

We pay special attention to single ALP emission in the $\eta^{(\prime)}\to\pi\pi a$ decay channel. This channel has been previously investigated in the context of the QCD axion using a U(3) chiral effective Lagrangian at leading order~\cite{Landini:2019eck}. 
In the framework of Resonance Chiral Theory, ref.\,\cite{Alves:2020xhf} went beyond leading order and included nonperturbative effects by means of the exchange of low-lying QCD scalar resonances. 
Existing literature also includes leading order calculations for a generic gluon-coupled ALPs~\cite{Gan:2020aco}.
Here, we go beyond existing leading order calculations and exploit the universality of the $\pi\pi$ final state interactions (FSI) via dispersion relations (see, {\it{e.g.}} \cite{Hanhart:2013vba,Holz:2015tcg,Stollenwerk:2011zz,Kubis:2015sga,Hanhart:2016pcd,Niecknig:2012sj,JPAC:2020umo,JPAC:2023nhq}) to account for the strong $\pi\pi$ final state rescattering in $\eta^{(\prime)}\to\pi\pi a$ decays.
We show that these corrections are significant, and, if neglected, can lead to a major underestimation of axio-hadronic decay rates, almost reaching an order of magnitude for certain regions of ALP parameter space.

We also consider the single ALP channel $\eta^{\prime}\to\eta\pi^0 a$, albeit with less care for quantifying the uncertainties stemming from the phase shift in the $\eta\pi^0$ final state rescattering.

Finally, we calculate, for the first time, the leading order $\eta^{(\prime)}$ decay rates involving multiple ALP emission, such as $\eta^{(\prime)}\to\pi^{0}aa$, $\eta^{\prime}\to\eta aa$ and $\eta^{(\prime)}\to aaa$, and provide a brief discussion of the magnitude of FSI corrections and the additional processes involved. We leave a complete treatment of FSI effects in multi-ALP decay channels to a future publication.

The structure of this article is as follows.
Sec.\,\ref{sec:formalism} outlines our theoretical framework, including our notation, leading order ALP-chiral Lagrangian, and derivation of the amplitudes for single ALP $\eta^{(\prime)}$ decays (Sec.\,\ref{sec:ALPLagrangian}); we end this section by specifying our conventions for decay kinematic variables (Sec.\,\ref{sec:kinematics}). The core contribution of this study is found in Sec.\,\ref{sec:rescattering}, where we use the framework of dispersion relations to extract $\pi\pi$ final state interactions effects using the $\pi\pi$ $S$-wave phase shift (Subsec.\,\ref{sec:phaseshift}).
In Sec.\,\ref{sec:pheno} we obtain the branching ratios of single ALP $\eta^{(\prime)}$ decays for two benchmark scenarios, and in Sec.\,\ref{sec:MultiALP} we obtain the leading order amplitudes and decay rates with multiple ALP emission. We summarize our main results and comment on experimental prospects in Sec.\,\ref{sec:summary}.

\section{Theoretical framework}
\label{sec:formalism}

\subsection{ALP-$\chi$PT Lagrangian and leading order decay amplitudes}
\label{sec:ALPLagrangian}

Generic ALP models show a diverse range of phenomenological signatures due to the many ways in which ALPs can couple to Standard Model particles. 
In this study, we shall concentrate on flavor-preserving ALP couplings to quarks and gluons, which control $\eta/\eta^{\prime}$ decays to final states containing hadrons and one or more ALPs. In particular, we will ignore more generic ALP interactions involving flavor-changing transitions, which are important in, e.g., Kaon and $B$-meson decays. We will also omit ALP couplings to photons and leptons; while they do not affect axio-hadronic $\eta/\eta^{\prime}$ decays, they are key parameters in determining the ALP's decay modes and lifetime, especially for ALP masses below the hadronic decay threshold, $m_a<3 m_\pi$. Generically, light ALPs produced in $\eta/\eta^{\prime}$ decays will either decay promptly to visible final states (e.g., $a\to \ell^+\ell^-$, $a\to\gamma\gamma$), or will travel a macroscopic distance before decaying, leading to events with displaced vertices or missing energy. When designing experimental ALP searches, one must combine our results for axio-hadronic $\eta/\eta^{\prime}$ decay rates with additional assumptions for the ALP decay modes and lifetime.

Several bases have been considered in the literature to express low energy ALP couplings. Here, we choose a basis in which the ALP couplings to quarks are exclusively expressed as Yukawa couplings (instead of derivative couplings). Not only this basis makes calculations more convenient, but this is often the more natural basis for UV completions of ALP models.
The Lagrangian in this {\it{Yukawa basis}} can be written as:
\begin{eqnarray}
\mathcal{L}^{Y}_{\rm{ALP}}~\supset~\frac{1}{2}\partial_{\mu}a\partial^{\mu}a-\frac{1}{2}\MPQ^{2}a^{2}-Q_{G}\frac{\alpha_{s}}{8\pi}\frac{a}{f_{a}}G\tilde{G}~+\sum_{q=u,d,s}m_{q}\bar{q}\left(e^{iQ_{q}\frac{a}{f_{a}}\gamma_{5}}\right)q.
\label{Eq:LagrangianALPqcdYukawa}
\end{eqnarray}	
Above, $\MPQ$ is a bare PQ-breaking contribution to the ALP mass; $f_{a}$ is the ALP decay constant; $G\tilde{G}\equiv \epsilon_{\mu\nu\alpha\beta}G_{\mu\nu}G^{\alpha\beta}$ with $G_{\mu\nu}$ denoting the gluon field strength tensor; and the PQ charges $Q_{G}$ and $Q_{q}$ parametrize the ALP couplings to gluons and quarks, respectively.
Note that (\ref{Eq:LagrangianALPqcdYukawa}) is defined at the GeV scale with the heavy quarks $c,b,t$ integrated out. 
Their PQ charges contribute implicitly to the ALP-gluon coupling through:
\begin{equation}
Q_{G}\equiv Q_{G}^{\rm{UV}}+Q_{c}+Q_{b}+Q_{t}\,.
\label{Eq:HeavyQuarksCoupling}
\end{equation}

By performing ALP-dependent chiral rotations of the quark fields, $q\to\exp(i\gamma_5 \frac{Q_{q}}{2}\frac{a}{f_{a}})q$, we can re-express the ALP model in \eqref{Eq:LagrangianALPqcdYukawa} in an equivalent basis in which the ALP couples only derivatively to fermions~\cite{Georgi:1986df}:
\begin{eqnarray}
\begin{aligned}
\mathcal{L}^{\partial}_{\rm{ALP}}&\supset-\left(Q_{G}^{\rm{UV}}+\sum_{q}Q_{q}\right)\frac{\alpha_{s}}{8\pi}\frac{a}{f_{a}}G\tilde{G}~+~\frac{\partial_{\mu}a}{f_{a}}\sum_{q=u,d,s}\frac{Q_{q}}{2}\bar{q}\gamma^{\mu}\gamma^{5}q\\
&+\sum_{i,j}\frac{g_{2}}{\sqrt{2}}\,V_{ij}^{\rm{CKM}}\,W_{\mu}^{+}\,\bar{u}_{L}^{i}\,e^{-i\left(\frac{Q_{u_{i}}-Q_{d_{j}}}{2}\frac{a}{f_{a}}\gamma_{5}\right)}\gamma^{\mu}\,d_{L}^{j}+\rm{h.c.}\end{aligned}
\label{Eq:LagrangianALPqcdDerivative}
\end{eqnarray}
The flavor-changing term in \eqref{Eq:LagrangianALPqcdDerivative} comes from chirally rotating the weak currents; it is often omitted in the literature, but it is necessary for the equivalency between \eqref{Eq:LagrangianALPqcdYukawa} and \eqref{Eq:LagrangianALPqcdDerivative} and for a consistent description of flavor-changing axionic meson decays \cite{Bauer:2021wjo}. Since in this study we are focused on flavor-conserving processes, this subtlety in basis-equivalency is not a concern to us. We shall now proceed with the Yukawa-basis representation of the ALP interactions in (\ref{Eq:LagrangianALPqcdYukawa}).

In order to extract the amplitudes for the decays $\eta/\eta^{\prime}\to\pi\pi a$ we need a framework to describe ALP interactions with mesons.
The first step is to map the QCD level Lagrangian $\mathcal{L}^{Y}_{\rm{ALP}}$ 
in (\ref{Eq:LagrangianALPqcdYukawa}) into Chiral Perturbation Theory ($\chi$PT)~\cite{Gasser:1984gg}.
Because we care about decays of the $\eta^{\prime}$ meson, we carry out our study relying on Large-$N_{C}$ $\chi$PT~\cite{Herrera-Siklody:1996tqr,Herrera-Siklody:1997pgy,Leutwyler:1997yr,Kaiser:2000gs}.
We recall that the large mass of the $\eta^{\prime}$ meson (which is composed predominantly of the pseudoscalar singlet $\eta_{0}$) comes from the anomalous breaking of the $\text{U}(1)_{A}$ global symmetry by the gluon density operator $G\tilde{G}$, which prevents the $\eta_{0}$ state to be a pseudo-Nambu-Goldstone boson (pNGB).
In the $N_{C}\to\infty$ limit, however, the $\text{U}(1)_{A}$ axial anomaly vanishes and the $\eta_{0}$---absent in the standard U(3) $\chi$PT---becomes a new light degree-of-freedom of the effective theory, {\it{i.e.}}, the ninth pNGB associated with the spontaneous symmetry breaking of $\text{U}(3)_{L}\times \text{U}(3)_{R}\to \text{U}(3)_{V}$.
In this regard, we map the ALP-gluon coupling term in (\ref{Eq:LagrangianALPqcdYukawa}) into $\chi$PT in a similar fashion as the $\eta_{0}$, following the heuristic prescription in \cite{tHooft:1986ooh}.

Within this framework, the leading order ALP-$\chi$PT Lagrangian can be written as~\cite{REDTOP:2022slw}:
\begin{eqnarray}\nonumber
\mathcal{L}_{\rm{ALP}}^{\chi\rm{PT@LO}}&=&~\frac{1}{2}\partial_{\mu}a\partial^{\mu}a-\frac{1}{2}\MPQ^{2}a^{2}-\frac{1}{2}m_{0}^{2}\left(\eta_{0}-\frac{Q_{G}}{\sqrt{6}}\frac{f_{\pi}}{f_{a}}a\right)^{2}\\
&&+~\frac{f_{\pi}^{2}}{4}{\rm{Tr}}\Big[\partial_{\mu}U^{\dagger}\partial^{\mu}U\Big]~+~\frac{f_{\pi}^{2}}{4}{\rm{Tr}}\Big[2B_{0}(M_{q}(a)U+M_{q}(a)^{\dagger}U^{\dagger})\Big]\,,
\label{Eq:Lag}
\end{eqnarray}
where $f_{\pi}=92.07$ MeV~\cite{Workman:2022ynf} is the pion decay constant; $B_{0}$ is a low energy constant related to the quark condensate, $\delta^{ij}B_{0}=-\langle q^{i}\bar{q}^{j}\rangle/f_{\pi}^{2}$; $m_{0}$ parametrizes the $\text{U}(1)_{A}$ anomaly contribution to the mass of the chiral singlet $\eta_{0}$; $M_{q}(a)$ is the ALP-dependent quark mass matrix,
\begin{equation}
M_{q}(a)\equiv\begin{pmatrix}
m_{u}e^{iQ_{u}a/f_{a}}&&\\
&m_{d}e^{iQ_{d}a/f_{a}}&\\
&&m_{s}e^{iQ_{s}a/f_{a}}&
\end{pmatrix}\,,
\end{equation}
and $U$ is the nonlinear representation of the pNGB chiral meson nonet,
\begin{equation}
U\equiv\exp{\left(\frac{i\sqrt{2}\Phi}{f_{\pi}}\right)}\,,
\end{equation}
with
\begin{eqnarray}
\Phi\equiv
\begin{pmatrix}
\frac{1}{\sqrt{2}}\pi_{3}+\frac{1}{\sqrt{6}}\eta_{8}+\frac{1}{\sqrt{3}}\eta_{0}&\pi^{+}&K^{+}\cr
\pi^{-}&-\frac{1}{\sqrt{2}}\pi_{3}+\frac{1}{\sqrt{6}}\eta_{8}+\frac{1}{\sqrt{3}}\eta_{0}&K^{0}\cr
K^{-}&\bar{K}^{0}&-\frac{2}{\sqrt{6}}\eta_{8}+\frac{1}{\sqrt{3}}\eta_{0}
\end{pmatrix}\,.
\label{matrix}
\end{eqnarray}

The quadratic mass term in (\ref{Eq:Lag}) mixes the ALP field $a$ with the neutral, strangeness-zero chiral mesons fields $\pi_{3},\eta_{8}$ and $\eta_{0}$.
We can express these quadratic Lagrangian terms as $\mathcal{L}_{\rm{ALP}}^{\chi\rm{PT@LO}}\supset-\frac{1}{2}\phi^{T}\,\mathds{M}_\phi^{2}\,\phi$ with $\phi\equiv\left(\pi_{3},\eta_{8},\eta_{0},a\right)$, and
\begin{equation}
\mathds{M}_\phi^{2}=
\begin{pmatrix}
\,\mu_{\pi_{3}}^{2}\,&~\mu_{\pi_{3}\eta_{8}}^{2}~&~\mu_{\pi_{3}\eta_{0}}^{2}~&~\mu_{a\pi}^{2}~\\
&\mu_{\eta_{8}}^{2}&\mu_{\eta_{8}\eta_{0}}^{2}&\mu_{a\eta_8}^{2}\\
&&\mu_{\eta_{0}}^{2}&\mu_{a\eta_0}^{2}\\
&&&\mu_{a}^{2}
\end{pmatrix}\,,
\label{Eq:massmatrix}
\end{equation}
where
\begin{eqnarray}
&\mu_{\pi_{3}}^{2}&=~B_{0}(m_{u}+m_{d}),\\
&\mu_{\pi_{3}\eta_{8}}^{2}&=~\frac{B_{0}}{\sqrt{3}}(m_{u}-m_{d}),\\
&\mu_{\pi_{3}\eta_{0}}^{2}&=~\sqrt{\frac{2}{3}}B_{0}(m_{u}-m_{d}),\\
&\mu_{\eta_{8}}^{2}&=~\frac{B_{0}}{3}(m_{u}+m_{d}+4m_{s}),\\
&\mu_{\eta_{8}\eta_{0}}^{2}&=~\frac{\sqrt{2}}{3}B_{0}(m_{u}+m_{d}-2m_{s}),\\
&\mu_{\eta_{0}}^{2}&=~m_{0}^{2}+\frac{2}{3}B_{0}(m_{u}+m_{d}+m_{s}),\label{Eq:eta0mass}\\
&\mu_{a\pi_3}^{2}&=~\frac{f_{\pi}}{f_{a}}B_{0}(m_{u}Q_{u}-m_{d}Q_{d}),\\
&\mu_{a\eta_8}^{2}&=~\frac{f_{\pi}}{f_{a}}\frac{B_{0}}{\sqrt{3}}(m_{u}Q_{u}+m_{d}Q_{d}-2m_{s}Q_{s}),\\
&\mu_{a\eta_0}^{2}&=~\frac{f_{\pi}}{f_{a}}\sqrt{\frac{2}{3}}B_{0}(m_{u}Q_{u}+m_{d}Q_{d}+m_{s}Q_{s})-\frac{f_{\pi}}{f_{a}}m_{0}^{2}\frac{Q_{G}}{\sqrt{6}},\\
&\mu_{a}^{2}&=~\frac{f_{\pi}^{2}}{f_{a}^{2}}B_{0}(m_{u}Q_{u}^{2}+m_{d}Q_{d}^{2}+m_{s}Q_{s}^{2})+\frac{f_{\pi}^{2}}{f_{a}^{2}}m_{0}^{2}\frac{Q_{G}^{2}}{6}+\MPQ^{2}.
\end{eqnarray}

In order to obtain the physical ALP state $a_{\rm{phys}}$ and meson states $\pi^{0},\eta$ and $\eta^{\prime}$, the mass matrix in (\ref{Eq:massmatrix}) needs to be diagonalized. This process can be done in two steps.
The first step is the removal of the ALP-meson mixing terms, parametrized by the angles $\theta_{a\pi},\theta_{a\eta_8},\theta_{a\eta_0}$. The second step is the diagonalization of the chiral meson mass terms. The resulting relations between the original and physical states are:
\begin{equation}\begin{pmatrix}
\pi_{3}\\
\eta_{8}\\
\eta_{0}\\
a
\end{pmatrix}=
\left(
    \begin{array}{ccccc}
        &  &   &\multicolumn{1}{: c}{\theta_{a\pi}}   \\
       &  \mathds{1}_{3\times3} &  & \multicolumn{1}{: c}{\theta_{a\eta_8}}   \\
        &   &  & \multicolumn{1}{: c}{\theta_{a\eta_0}}   \\
      \cdashline{1-4}
       -\theta_{a\pi}&-\theta_{a\eta_8}&-\theta_{a\eta_0}& \multicolumn{1}{: c}{1} 
    \end{array}
  \right)
  \left(
    \begin{array}{ccccc}
        &  &   &\multicolumn{1}{: c}{0}   \\
       &  \mathds{R}_{3\times3} &  & \multicolumn{1}{: c}{0}   \\
        &   &  & \multicolumn{1}{: c}{0}   \\
      \cdashline{1-4}
       ~0~&~0~&~0~& \multicolumn{1}{: c}{~1~} 
    \end{array}
  \right)
  \begin{pmatrix}
\pi^{0}\\
\eta\\
\eta^{\prime}\\
a_{\rm{phys}}
\end{pmatrix}
\,,
\label{Eq:Mixing}
\end{equation}
where $\mathds{R}$ is an orthogonal matrix that diagonalizes of the $\pi^0$-$\eta$-$\eta^\prime$ meson subsystem, given by~\cite{Leutwyler:1996np,Escribano:2016ntp}:
\begin{equation}
\mathds{R}=\begin{pmatrix}
1&-\theta_{\pi\eta}&-\theta_{\pi\eta^{\prime}}\\
\left(\theta_{\pi\eta}\cos\theta_{\eta\eta^{\prime}}+\theta_{\pi\eta^{\prime}}\sin\theta_{\eta\eta^{\prime}}\right)&\cos\theta_{\eta\eta^{\prime}}&\sin\theta_{\eta\eta^{\prime}}\\
\left(\theta_{\pi\eta^{\prime}}\cos\theta_{\eta\eta^{\prime}}-\theta_{\pi\eta}\sin\theta_{\eta\eta^{\prime}}\right)&-\sin\theta_{\eta\eta^{\prime}}&\cos\theta_{\eta\eta^{\prime}}\\
\end{pmatrix}
\,.
\end{equation}
Above, $\theta_{\eta\eta^{\prime}}$, $\theta_{\pi\eta}$, and $\theta_{\pi\eta^{\prime}}$ are, respectively, the $\eta_{8}$-$\eta_{0}$, $\pi_{3}$-$\eta_{8}$, and $\pi_{3}$-$\eta_{0}$ mixing angles. 

For our phenomenological analysis in Sec.~\ref{sec:pheno}, we will use the $\eta$-$\eta^{\prime}$ mixing angle measured by KLOE, and also set $\theta_{\eta\eta^{\prime}}=-13.3(5)^{\circ}$~\cite{KLOE:2006guu}, $m_{0}=1.03$ GeV~\cite{Guo:2015xva}, and $\theta_{\pi\eta}=0.018$, $\theta_{\pi\eta^{\prime}}=0.0049$~\cite{Gao:2022xqz}. This choice results in leading order physical masses of $m_{\eta}^{(0)}=534$ MeV  for the $\eta$ meson and $m_{\eta^{\prime}}^{(0)}=1.14$ GeV for the $\eta^{\prime}$ meson, which disagree significantly with the experimental values of $m_{\eta}=548$\;MeV and $m_{\eta^{\prime}}=958$\;MeV. Resolving this discrepancy requires one to go to next-to-leading order in the $\chi$PT expansion \cite{Georgi:1993jn}, which is beyond the scope of this work. In what follows we will adopt this numerical choice of parameters, but use the experimental values for the $\eta$ and $\eta^\prime$ masses when computing their decay rates.

It is relatively straightforward to obtain the ALP-meson mixing terms $\theta_{a\pi},\theta_{a\eta_8},\theta_{a\eta_0}$ in (\ref{Eq:Mixing})\footnote{A simple trick is to integrate out the $\pi_3$, $\eta_8$ and $\eta_0$ states via the equations of motion.}. Working first in the PQ-preserving limit $\MPQ=0$ (in which case the ALP can be identified with the QCD axion), the mass-squared of the physical axion $a_{\rm{phys}}$ is given by
\begin{equation}
\big(m_a^{\text{\tiny (PQ)}}\big)^2=\left(Q_{u}+Q_{d}+Q_{s}+Q_{G}\right)^{2}\frac{B_{0}m_{u}m_{d}m_{s}}{\left(m_{u}m_{d}+m_{u}m_{s}+m_{d}m_{s}+\frac{6B_{0}m_{u}m_{d}m_{s}}{m_{0}^{2}}\right)}\frac{f_{\pi}^{2}}{f_{a}^{2}}\,,\label{Eq:ALPmassPQ}
\end{equation}
and the axion-meson mixing angles are given by
\begin{eqnarray}
\theta_{a\pi}^{^{\text{\tiny (PQ)}}}&=&-\frac{f_{\pi}}{f_{a}}\frac{1}{(1+\epsilon)}\bigg(\frac{Q_{u}m_{u}-Q_dm_d}{m_{u}+m_{d}}+\frac{m_{u}-m_{d}}{m_{u}+m_{d}}\;\frac{Q_{s}+Q_{G}}{2}+\epsilon\,\frac{Q_{u}-Q_{d}}{2}\bigg),
\label{mixing1}\\
\nonumber\\
\theta_{a\eta_8}^{^{\text{\tiny (PQ)}}}&=&\frac{f_{\pi}}{f_{a}}\frac{\sqrt{3}}{2}\frac{1}{(1+\epsilon)}\Bigg(Q_{s}+\frac{Q_{G}}{3}-\epsilon\frac{(Q_{u}+Q_{d}+2Q_{G}/3)+\frac{2B_{0}m_{s}}{m_{0}^{2}}(Q_{u}+Q_{d}-2Q_{s})}{1+\frac{6B_{0}m_{s}}{m_{0}^{2}}}\Bigg),~~~~~~~~\label{mixing2}\\
\nonumber\\
\theta_{a\eta_0}^{^{\text{\tiny (PQ)}}}&=&\frac{f_{\pi}}{f_{a}}\frac{1}{\sqrt{6}}\frac{1}{(1+\epsilon)}\Bigg(Q_{G}+\epsilon\frac{Q_{G}-\frac{6B_{0}m_{s}}{m_{0}^{2}}(Q_{u}+Q_{d}+Q_{s})}{1+\frac{6B_{0}m_{s}}{m_{0}^{2}}}\Bigg),\label{mixing3}
\end{eqnarray}
where the small parameter $\epsilon$, of order $\mathcal{O}(m_\pi^2/m_K^2)$, is given by
\begin{equation}
\epsilon\equiv\frac{m_{u}m_{d}}{m_{s}(m_{u}+m_{d})}\left(1+6\,\frac{B_{0} m_{s}}{m_{0}^{2}}\right)\approx 0.04\,.
\end{equation}

Next, we can consider the PQ-breaking case with a finite $\MPQ$ mass (when the ALP is no longer the QCD axion). Here, we assume that $f_a\gg f_\pi$ and neglect terms of $\mathcal{O}(\theta^2_{a\phi})$, where $\phi=\pi_3,\,\eta_8,\,\eta_0$. In this case, the mass-squared of the physical ALP $a_{\rm{phys}}$ is:
\begin{equation}
m_a^{2}=\big(m_a^{\text{\tiny (PQ)}}\big)^2+\MPQ^{2}\,.
\label{Eq:ALPmass}
\end{equation}
Neglecting $\pi$-$\eta$ and $\pi$-$\eta^\prime$ mixing (i.e., making the approximation of $\theta_{\pi\eta^{(\prime)}}\to 0$), the ALP-meson mixing angles are given by
\begin{subequations}
\begin{alignat}{2}
\theta_{a\pi} =&~ \theta_{a\pi}^{^\text{\tiny (PQ)}}\Bigg(1\,+\,\frac{\MPQ^2}{m_\pi^2-m_a^{2}}\Bigg)\,,\label{Eq:ALPpionMixingAngle}\\
\vspace{2em}
\theta_{a\eta_8} =&~ \theta_{a\eta_8}^{^\text{\tiny (PQ)}}\Bigg(1\,+\,\cos^2\!\theta_{\eta\eta^\prime}\,\frac{\MPQ^2}{\,m_\eta^2-m_a^{2}\,}\,+\,\sin^2\!\theta_{\eta\eta^\prime}\,\frac{\MPQ^2}{\,m_{\eta^\prime}^2-m_a^{2}\,}\Bigg)\label{Eq:ALPeta8MixingAngle}\\
&+\theta_{a\eta_0}^{^\text{\tiny (PQ)}}\,\frac{\sin2\theta_{\eta\eta^\prime}}{2}\Bigg(\frac{\MPQ^2}{\,m_{\eta^\prime}^2-m_a^{2}\,}\,-\,\frac{\MPQ^2}{\,m_{\eta}^2-m_a^{2}\,}\Bigg)\,,\nonumber\\
\vspace{2em}
\theta_{a\eta_0} =&~ \theta_{a\eta_0}^{^\text{\tiny (PQ)}}\Bigg(1\,+\,\sin^2\!\theta_{\eta\eta^\prime}\,\frac{\MPQ^2}{\,m_\eta^2-m_a^{2}\,}\,+\,\cos^2\!\theta_{\eta\eta^\prime}\,\frac{\MPQ^2}{\,m_{\eta^\prime}^2-m_a^{2}\,}\Bigg)\label{Eq:ALPeta0MixingAngle}\\
&+\theta_{a\eta_8}^{^\text{\tiny (PQ)}}\,\frac{\sin2\theta_{\eta\eta^\prime}}{2}\Bigg(\frac{\MPQ^2}{\,m_{\eta^\prime}^2-m_a^{2}\,}\,-\,\frac{\MPQ^2}{\,m_{\eta}^2-m_a^{2}\,}\Bigg)\,.\nonumber
\end{alignat}
\end{subequations}
Note that the expressions above are only valid in the small mixing angle approximation, i.e., when $\theta_{a\pi}^{^\text{\tiny (PQ)}},\theta_{a\eta_8}^{^\text{\tiny (PQ)}},\theta_{a\eta_0}^{^\text{\tiny (PQ)}}\ll1$; $\MPQ^2/|m_\pi^2-m_a^2|\ll f_a/f_\pi$ for (\ref{Eq:ALPpionMixingAngle}); and $\MPQ^2/|m_{\eta^{(\prime)}}^2-m_a^2|\ll f_a/f_\pi$ for (\ref{Eq:ALPeta8MixingAngle}) and (\ref{Eq:ALPeta0MixingAngle}).

We are now in a position to obtain the leading order amplitudes\footnote{In order to lighten the notation, we shall suppress from this point forward the subscript in the physical ALP state, i.e., we will take $a_{\rm{phys}}\to a$.} for $\eta^{(\prime)}\to\pi^{0}\pi^{0}a$, $\eta^{(\prime)}\to\pi^{+}\pi^{-}a$, and $\eta^{\prime}\to\eta\pi^{0}a$.
These follow from the ALP Lagrangian in (\ref{Eq:Lag}) with the field redefinitions from (\ref{Eq:Mixing}):
\begin{eqnarray}
\mathcal{A}(\eta\to\pi^{0}\pi^0 a)&=&~\frac{m_\pi^2}{f_\pi^2}\,\Bigg[C_{\eta}\,\frac{(m_uA_u+m_dA_d)}{m_u+m_d}\label{Eq:AmplitudeEtaNeutral}\\
&&~~~~~~~~+\bigg(2\theta_{\pi\eta^{\prime}}C_{\eta}C_{\eta^\prime}+\theta_{\pi\eta}(C_{\eta}^2-C_{\eta^\prime}^2)\bigg)\,\frac{(m_uA_u-m_dA_d)}{m_u+m_d}\Bigg]\,,\nonumber\\
\nonumber\\
\mathcal{A}(\eta^{\prime}\to\pi^0\pi^{0}a)&=&~\frac{m_\pi^2}{f_\pi^2}\,\Bigg[C_{\eta^\prime}\,\frac{(m_uA_u+m_dA_d)}{m_u+m_d}\label{Eq:AmplitudeEtaPrimeNeutral}\\
&&~~~~~~~~+\bigg(2\theta_{\pi\eta}C_{\eta}C_{\eta^\prime}+\theta_{\pi\eta^\prime}(C_{\eta^\prime}^2-C_{\eta}^2)\bigg)\,\frac{(m_uA_u-m_dA_d)}{m_u+m_d}\Bigg]\,,\nonumber
\\
\nonumber\\
\mathcal{A}(\eta\to\pi^{+}\pi^{-}a)&=&~\frac{m_\pi^2}{f_\pi^2}\,\Bigg[C_{\eta}\,\frac{(m_uA_u+m_dA_d-2(m_u-m_d)\theta_{a\pi}/3)}{m_u+m_d}\label{Eq:AmplitudeEtaCharged}\\
&&~~~~~~~~-\theta_{\pi\eta}\,\frac{(m_uA_u-m_dA_d)}{m_u+m_d}\Bigg]
-\theta_{\pi\eta}\frac{\theta_{a\pi}}{f_{\pi}^{2}}\Bigg[s-\,\frac{m_{\eta}^{2}+2m_{\pi}^{2}+m_{a}^{2}}{3}\Bigg]\,,\nonumber\\
\nonumber
\end{eqnarray}
\begin{eqnarray}
\mathcal{A}(\eta^{\prime}\to\pi^{+}\pi^{-}a)&=&~\frac{m_\pi^2}{f_\pi^2}\,\Bigg[C_{\eta^\prime}\,\frac{(m_uA_u+m_dA_d-2(m_u-m_d)\theta_{a\pi}/3)}{m_u+m_d}\label{Eq:AmplitudeEtaPrimeCharged}\\
&&~~~~~~~~-\theta_{\pi\eta^\prime}\,\frac{(m_uA_u-m_dA_d)}{m_u+m_d}\Bigg]
-\theta_{\pi\eta^\prime}\frac{\theta_{a\pi}}{f_{\pi}^{2}}\Bigg[s-\,\frac{m_{\eta^\prime}^{2}+2m_{\pi}^{2}+m_{a}^{2}}{3}\Bigg]\,,\nonumber\\
\nonumber\\
\mathcal{A}(\eta^{\prime}\to\eta\pi^{0}a)&=&~\frac{m_\pi^2}{f_\pi^2}\,\Bigg[C_{\eta}C_{\eta^\prime}\,\frac{(m_uA_u-m_dA_d)}{m_u+m_d}\label{Eq:AmplitudeEtaPrimetoEta}\\
&&~~~~~~~~-\theta_{\pi\eta}C_{\eta^\prime}^2\;\frac{(m_uA_u+m_dA_d)C_{\eta^\prime}-4m_sA_sC_\eta}{m_u+m_d}\nonumber\\
&&~~~~~~~~-\theta_{\pi\eta^\prime}C_{\eta}^2\;\frac{(m_uA_u+m_dA_d)C_{\eta}+4m_sA_sC_{\eta^\prime}}{m_u+m_d}\Bigg]\,,\nonumber
\end{eqnarray}
where $m_{\pi}^{2}\equiv B_{0}(m_{u}+m_{d})$ is the leading order pion mass in the isospin limit; the Mandelstam variable $s$ is defined as $s\equiv(p_{\eta}-p_{a})^{2}$ (see Subsec.\,\ref{sec:kinematics}); and
\begin{eqnarray}\label{etapipiALPquartic}
C_{\eta}\,&\equiv&\,\frac{\cos\!\theta_{\eta\eta^\prime}}{\sqrt{3}}-\frac{\sin\!\theta_{\eta\eta^\prime}}{\sqrt{3/2}},\label{Ceta}\\
C_{\eta^\prime}\,&\equiv&\,\frac{\cos\!\theta_{\eta\eta^\prime}}{\sqrt{3/2}}+\frac{\sin\!\theta_{\eta\eta^\prime}}{\sqrt{3}},\label{CetaP}\\
A_u\,&\equiv&\frac{f_\pi}{f_a}Q_u+\theta_{a\pi}+\frac{\theta_{a\eta_8}}{\sqrt{3}}+\frac{\theta_{a\eta_0}}{\sqrt{3/2}},\label{Au}\\
A_d\,&\equiv&\frac{f_\pi}{f_a}Q_d-\theta_{a\pi}+\frac{\theta_{a\eta_8}}{\sqrt{3}}+\frac{\theta_{a\eta_0}}{\sqrt{3/2}},\label{Ad}\\
A_s\,&\equiv&\frac{f_\pi}{f_a}\frac{Q_s}{\sqrt{2}}-\frac{\theta_{a\eta_8}}{\sqrt{3/2}}+\frac{\theta_{a\eta_0}}{\sqrt{3}}.\label{As}
\end{eqnarray}
The amplitudes in \eqref{Eq:AmplitudeEtaNeutral}–\eqref{Eq:AmplitudeEtaPrimetoEta} scale as $\sim1/f_{a}$ and receive contributions from a direct $a\pi\pi\eta^{(\prime)}$ quartic term in the leading order Lagrangian, as well as from mixing terms stemming from quartic SM meson interactions where one meson state mixes with the ALP\footnote{See Refs.~\cite{Alves:2020xhf,Gan:2020aco,REDTOP:2022slw} for similar calculations of these amplitudes, and Ref.~\cite{Aloni:2018vki} for a calculation of the related amplitude $a\to\pi\pi\eta^{(\prime)}$.}. Note also that subleading terms involving $\pi-\eta^{(\prime)}$ mixing (i.e., proportional to $\theta_{\pi\eta^{(\prime)}}$) generally provide only percent-level corrections to the total amplitudes in \eqref{Eq:AmplitudeEtaNeutral}–\eqref{Eq:AmplitudeEtaPrimeCharged}, and in most cases can be neglected. The one instance in which the $\theta_{\pi\eta^{(\prime)}}$-suppressed contributions are important is in the KSVZ QCD axion limit of \eqref{Eq:AmplitudeEtaPrimetoEta} (i.e., with $\MPQ=0$, $Q_q=0$, and finite $Q_G^{\rm{UV}}$). In this case, the first term in \eqref{Eq:AmplitudeEtaPrimetoEta} vanishes identically (cf.\,(\ref{mixing1}), (\ref{mixing2}) and (\ref{mixing3})), leaving the $\theta_{\pi\eta^{(\prime)}}$-suppressed terms as the dominant leading order contributions to $\mathcal{A}(\eta^{\prime}\to\eta\pi^{0}a)$.

\subsection{Decay kinematics}
\label{sec:kinematics}

Before moving on to the main section of this paper, we briefly outline our conventions for kinematic variables. We define the amplitude for the decay $\eta\to\pi\pi a$ in the usual way,
\begin{equation}
\langle\pi^{i}(p_{1})\pi^{j}(p_{2})a(p_{a})|T|\eta(p_{\eta})\rangle=(2\pi)^{4}\delta^{4}(p_{\eta}-p_{1}-p_{2}-p_{a})\delta^{ij}\mathcal{M}(s,t,u)\,,
\label{Eq:Amplitude}
\end{equation}
where $i,j$ refer to the isospin indices of the pion.
In the following, we will consider both the neutral $\eta\to\pi^{0}\pi^{0} a$ and charged $\eta\to\pi^{+}\pi^{-} a$ decay channels, which only differ by isospin-breaking effects.
The Mandelstam variables for the decay process are given by:
\begin{eqnarray}
s&=&(p_{\eta}-p_{a})^{2}=(p_{1}+p_{2})^{2}\,,\\
t&=&(p_{\eta}-p_{1})^{2}=(p_{2}+p_{a})^{2}\,,\\
u&=&(p_{\eta}-p_{2})^{2}=(p_{1}+p_{a})^{2}\,,
\end{eqnarray}
which fulfill the relation
\begin{equation}
s+t+u=m_{\eta}^{2}+2m_{\pi}^{2}+m_{a}^{2}\,.
\end{equation}
In the center-of-mass system of the two pions, $t$ and $u$ can be expressed in terms of $s$ and $\theta_{s}$ according to
\begin{eqnarray}
t(s,\cos\theta_{s})&=&\frac{1}{2}\left(m_{\eta}^{2}+2m_{\pi}^{2}+m_{a}^{2}-s+\kappa_{\pi\pi}(s)\cos\theta_{s}\right)\,,\\
u(s,\cos\theta_{s})&=&\frac{1}{2}\left(m_{\eta}^{2}+2m_{\pi}^{2}+m_{a}^{2}-s-\kappa_{\pi\pi}(s)\cos\theta_{s}\right)\,,
\end{eqnarray}
where $\cos\theta_{s}$ refers to the scattering angle,
\begin{equation}
\cos\theta_{s}\equiv\frac{t-u}{\kappa_{\pi\pi}(s)}\,,\quad \kappa_{\pi\pi}(s)\equiv\sigma_{\pi}(s)\lambda^{1/2}(m_{\eta}^{2},m_{a}^{2},s)\,,
\end{equation}
with $\sigma_{\pi}(s)\equiv\sqrt{1-4m_{\pi}^{2}/s}$, and the K\"{a}llen function $\lambda(a,b,c)\equiv a^{2}+b^{2}+c^{2}-2(ab+ac+bc)$.

The partial decay rate is given by~\cite{Workman:2022ynf}
\begin{equation}
\Gamma(\eta\to\pi\pi a)=\frac{1}{S_{\pi_{1}\pi_{2}}}\frac{1}{(2\pi)^{3}}\frac{1}{32m_{\eta}^{3}}\int ds\,dt\,|\mathcal{M}(s,t,u)|^{2}\,,
\label{Eq:DecayWidth}
\end{equation}
where $S_{\pi_{1}\pi_{2}}$ is a combinatorial factor that accounts for the number of identical particles in the final state; in particular, $S_{\pi^{+}\pi^{-}}=1$ and $S_{\pi^{0}\pi^{0}}=2!$. 
The boundaries of the physical decay region in $t$ lie within the interval $[t_{\rm{min}}(s),t_{\rm{max}}(s)]$, with
\begin{eqnarray}\label{tDalitz}
t_{\rm{max/min}}(s)=\frac{1}{2}\Bigg[m_{\eta}^{2}+m_{a}^{2}+2m_{\pi}^{2}-s\pm\frac{\lambda^{1/2}(s,m_{\eta}^{2},m_{a}^{2})\lambda^{1/2}(s,m_{\pi}^{2},m_{\pi}^{2})}{s}\Bigg]\,,
\label{tmaxmin}
\end{eqnarray}
while the allowed range for $s$ is
\begin{eqnarray}\label{sDalitz}
s_{\rm{min}}=4m_{\pi}^{2}\,,\quad s_{\rm{max}}=(m_{\eta}-m_{a})^{2}\,.
\end{eqnarray}
The region defined by \eqref{tDalitz} and \eqref{sDalitz} defines the Dalitz phase space.

The derivation of the decay rate for the analogous reaction $\eta^{\prime}\to\pi\pi a$ is formally identical to that of (\ref{Eq:DecayWidth}) with the appropriate replacements, {\it{e.g.}}, $m_{\eta}\to m_{\eta^{\prime}}$.

\section{Pion-pion final state interactions with dispersion relations}\label{sec:rescattering}

In any process with hadrons in the final state, final state interactions (FSI) from strong rescattering effects must be taken into account if one aims to make predictions with a reasonable degree of accuracy. 

A model independent approach that ressums FSI is given by dispersion theory.
Based on the fundamental physical principles of unitarity and analyticity, dispersion relations determine the amplitude up to certain subtraction constants that can be fixed by matching to the results of the effective theory, or from fits to experimental data, when available.

Because the ressummation of the FSI is imposed by construction, it becomes specially relevant and useful where perturbative, EFT-based approaches, such as $\chi$PT, are doomed to fail. This is generically the case of low energy hadronic processes involving low-lying QCD resonances in the physical region, such as the SM decays $\eta/\eta^{\prime}\to3\pi$~\cite{Colangelo:2018jxw} and $\eta^{\prime}\to\eta\pi\pi$~\cite{Isken:2017dkw}, as well as the BSM decays $\eta/\eta^{\prime}\to\pi\pi a$ we wish to consider here. In particular, the final state pions in these processes undergo strong FSI that could significantly perturb the spectrum, most notably in the isospin-zero $S$-wave channel with the corresponding appearance of the $\sigma$ meson resonance.

Alternative approaches to the treatment of FSI are unitarized versions of $\chi$PT, such as the Inverse Amplitude Method~\cite{Truong:1988zp,Dobado:1989qm,Truong:1991gv,Dobado:1996ps} or the $N/D$ method~\cite{Chew:1960iv,Oller:1998zr,Oller:2000ma}. However, in addition to the uncertainties in the determination of the low energy constants of the $\mathcal{O}(p^{4})$ $\chi$PT Lagrangian, these methods yield amplitudes that may not satisfy exact unitarity\footnote{These violations of unitarity are generically small (see, {\it{e.g.}}, refs.~\cite{Guerrero:1998ei,Gonzalez-Solis:2018xnw}).}~\cite{Badalian:1981xj,GomezNicola:2001as,Xiao:2001pt,Ledwig:2014cla}.
We therefore opt for the dispersive method over these alternative approaches for our analysis.
In the following, we present a detailed application of dispersion relations to the axio-hadronic $\eta/\eta^{\prime}\to\pi\pi a$ decays.

Using basic theorems of complex analysis, dispersion relations connect the discontinuity of the amplitude along its branch cut to the function itself via an integral equation that can be solved in terms of the so-called Omn\`{e}s equation, whose solution is given by the well-known $\pi\pi$ scattering phase shift extracted from the Roy equations~\cite{Roy:1971tc}. 
In our analysis, we include the $\pi\pi$ FSI stemming from the direct $s$-channel and neglect crossed-channel contributions due to the lack of information on $\pi a$ scattering; we shall return to this point below.

The decay amplitude $\mathcal{M}(s,t,u)$ in (\ref{Eq:Amplitude}) can be decomposed as
\begin{equation}
\mathcal{M}(s,t,u)=\sum_{\ell=0}^{\infty}(2\ell+1)P_{\ell}(\cos\theta_{s})m_{\ell}(s)\,,
\label{Eq:PartialWaveDecomposition}
\end{equation}
where $P_{\ell}(\cos\theta_{s})$ are the Legendre polynomials and $m_{\ell}(s)$ are the partial waves of angular momentum $\ell$, which can be obtained from the full amplitude by inversion of (\ref{Eq:PartialWaveDecomposition})
\begin{equation}
m_{\ell}(s)=\frac{1}{2}\int_{-1}^{1}d(\cos\theta_{s})P_{\ell}(\cos\theta_{s})\mathcal{M}(s,t(s,\cos\theta_{s}),u(s,\cos\theta_{s}))\,.
\label{Eq:PartialWaves}
\end{equation}

In practice, one can substitute the infinite sum of partial waves in the $s$-channel in (\ref{Eq:PartialWaveDecomposition}) by three finite sums of one Mandelstam variable only~\cite{JPAC:2020umo,JPAC:2021rxu}.
Explicitly,
\begin{eqnarray}
\mathcal{M}(s,t,u)&=&\sum_{\ell=0}^{\ell_{\rm{max}}}(2\ell+1)P_{\ell}(\cos\theta_{s})M_{\ell}(s)\nonumber\\
&+&\sum_{\ell=0}^{\ell_{\rm{max}}}(2\ell+1)P_{\ell}(\cos\theta_{t})M_{\ell}(t)+\sum_{\ell=0}^{\ell_{\rm{max}}}(2\ell+1)P_{\ell}(\cos\theta_{u})M_{\ell}(u)\,.
\label{Eq:PartialWaveDecomposition2}
\end{eqnarray}

Isospin conservation in the decay $\eta^{(\prime)}\to\pi\pi a$ constrains the total isospin of the final state pairs $\pi\pi$ and $\pi a$ to $I=0$ and $I=1$, respectively.
Consequently, only even partial waves contribute.
Given the smallness of the available phase space we truncate each sum in (\ref{Eq:PartialWaveDecomposition2}) at $\ell_{\rm{max}}=0$, so that only the $S$-wave contributes and the decomposition of the amplitude becomes
\begin{eqnarray}
\mathcal{M}(s,t,u)=M_{0}(s)+M_{0}(t)+M_{0}(u)\,.
\label{Eq:PartialWaveDecomposition3}
\end{eqnarray}
Projecting (\ref{Eq:PartialWaveDecomposition3}) into partial waves through (\ref{Eq:PartialWaves}) we obtain:
\begin{equation}
m_{0}(s)=M_{0}(s)+\hat{M}_{0}(s)\,,
\label{Eq:Swave}
\end{equation}
where $\hat{M}_{0}(s)$ is the so-called inhomogeneity, which is given by:
\begin{equation}
\hat{M}_{0}(s)=\frac{1}{2}\int_{-1}^{1}d(\cos\theta_{s})\left[M_{0}(t(s,\cos\theta_{s}))+M_{0}(u(s,\cos\theta_{s}))\right]\,.
\label{Eq:Inhomogeneity}
\end{equation}
The structure of (\ref{Eq:Swave}) is thus clear: the first term, $M_{0}(s)$, contains the right-hand cut contribution and accounts for $s$-channel $\pi\pi$ rescattering, while the second one, $\hat{M}_{0}(s)$, represents the $s$-channel projection of the left-hand cut contributions due to the $t$-and $u$-channels. One could model the left-hand cut contributions by the exchange of additional low-lying scalar resonances via Resonance Chiral Theory, such as the $a_{0}(980)$; however, these contributions would depend strongly on the coupling of $a_{0}(980)$ to the chiral mesons~\cite{Alves:2020xhf}.
As we do not have information on $\pi a$ scattering at our disposal, we simplify (\ref{Eq:Swave}) by neglecting $\hat{M}_{0}(s)$.

\begin{figure}
\centering\includegraphics[width=0.35\textwidth]{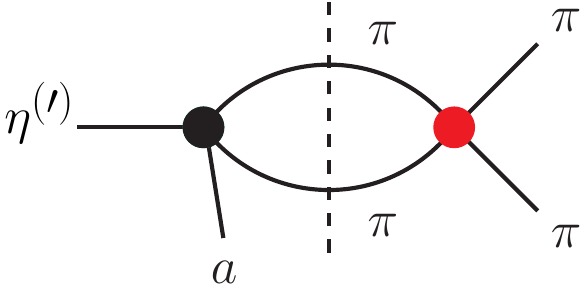}
\caption{Schematic representation of the $\pi\pi$ contribution to the discontinuity of the partial wave $M_{0}(s)$ (cf.~(\ref{Eq:Discontinuity})).
The black circle represents the $s$-channel $S$-wave projection of the $\eta^{(\prime)}\to\pi\pi a$ decay amplitude, while the red circle represents the $S$-wave $\pi\pi$ scattering amplitude.}
\label{Fig:Discontinuity} 
\end{figure}

To apply dispersion relations to the $M_{0}(s)$ function, one writes its unitarity relation as (see Fig.~\ref{Fig:Discontinuity} for a diagrammatic interpretation):
\begin{equation}
{\rm{disc}}M_{0}(s)=2iM_{0}(s)\sin\delta_{0}^{0}(s)e^{-i\delta_{0}^{0}(s)}\,,
\label{Eq:Discontinuity}
\end{equation}
where the function $\delta_{0}^{0}(s)$ is the $\pi\pi$ $S$-wave scattering phase shift; below the first inelastic threshold (in this case, the $K\bar{K}$ threshold) the phase of the partial-wave equals $\delta_{0}^{0}(s)$, as required by Watson's final state interaction theorem~\cite{Watson:1952ji,Watson:1954uc}.
Given the discontinuity relation in (\ref{Eq:Discontinuity}), one can write an unsubtracted dispersion relation for $M_{0}(s)$ as:
\begin{equation}
M_{0}(s)=\frac{1}{2\pi i}\int_{4m_{\pi}^{2}}^{\infty}ds^{\prime}\frac{{\rm{disc}}M_{0}(s^{\prime})}{s^{\prime}-s}\,,
\label{Eq:DR}
\end{equation}
which can be solved in terms of the usual Omn\`{e}s function~\cite{Omnes:1958hv}
\begin{equation}
    \Omega_{0}^{0}(s)=\exp\left[\frac{s}{\pi}\int_{4m_{\pi}^{2}}^{\infty}\frac{ds^{\prime}}{s^{\prime}}\frac{\delta_{0}^{0}(s^{\prime})}{s^{\prime}-s}\right]\,.
\label{Eq:Omnes}
\end{equation}
The most general solution of (\ref{Eq:DR}) can be written as
\begin{equation}
M_{0}(s)=P(s)\Omega_{0}^{0}(s)\,,
\label{Eq:GeneralOmnes}
\end{equation}
where the polynomial $P(s)$ is a real subtraction function not directly related to $\pi\pi$ rescattering.
At low energies, the dispersive amplitude $M_{0}(s)$ can be matched to the chiral one.
Performing the matching in the limit of no rescattering, {\it{i.e.}}, $\delta_{0}^{0}(s)\to0$, we have $\Omega_{0}^{0}(s)\to1$, so that the subtraction polynomial $P(s)$ in (\ref{Eq:GeneralOmnes}) can be identified exactly with the leading order expressions given in (\ref{Eq:AmplitudeEtaNeutral}), (\ref{Eq:AmplitudeEtaCharged}), (\ref{Eq:AmplitudeEtaPrimeNeutral}) and (\ref{Eq:AmplitudeEtaPrimeCharged}).

\subsection{Phase shift input and Omn\`{e}s function}\label{sec:phaseshift}

The key input in the calculation of the Omn\`{e}s equation in (\ref{Eq:Omnes}) is the $\pi\pi$ phase shift $\delta_{0}^{0}(s)$.
For the decay $\eta\to\pi\pi a$ (as well as for $\eta^{\prime}\to\pi\pi a$), the invariant mass of the pion-pion system is well below the first inelastic threshold, namely, the $K\bar{K}$ threshold.
Therefore, the single-channel dispersion relation correctly describes the final state interactions, with no need to consider multichannel rescattering effects explicitly, such as those from $K\bar{K}$ intermediate states.
For our analysis, we employ the parametrization for $\delta_{0}^{0}(s)$ resulting from the solution to the Roy equations~\cite{Garcia-Martin:2011iqs}, further imposing the asymptotic condition of $\delta_{0}^{0}(s)\to\pi$ when $s\to\infty$; this condition is reasonable since it is roughly satisfied at the $K\bar{K}$ threshold.
Note that this asymptotic behavior of the phase shift implies that the Omn\`{e}s function $\Omega_{0}^{0}(s)$ falls off as $1/s$, as expected.
Given our elastic approximation, we take $\delta_{0}^{0}(s)$ to be constant (or nearly constant) in the inelastic region above the $K\bar{K}$ threshold. This can be implemented by smoothly guiding the phase to $\pi$ above $\Lambda^{2}=4m_{K}^{2}$ through~\cite{Moussallam:1999aq}
\begin{eqnarray}
\delta_{0}^{0}(s\ge\Lambda^{2})=\pi+\left(\delta_{0}^{0}(\Lambda^{2})-\pi\right)f\left(\frac{s}{\Lambda^{2}}\right)\,,\quad f(x)=\frac{2}{1+x^{3/2}}\,.
\label{Eq:PhaseContinuation}
\end{eqnarray}

\begin{figure}
\centering\includegraphics[width=0.75\textwidth]{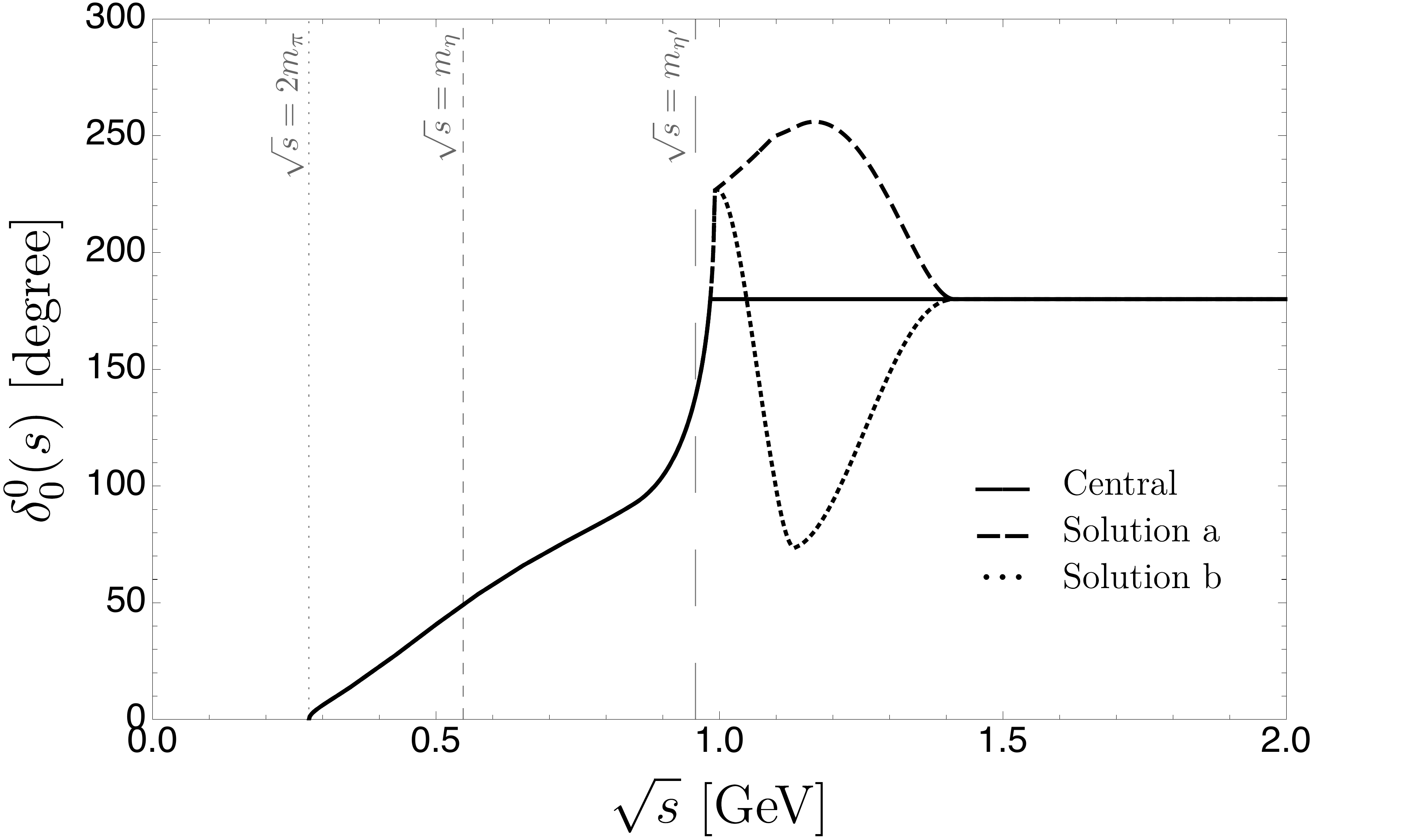}
\caption{The three examples of phase shift inputs discussed in the text. 
The vertical lines indicate the phase space boundaries for the decays $\eta\to\pi\pi a$ ($2m_{\pi}\leq\sqrt{s}\leq m_{\eta}$) and $\eta^{\prime}\to\pi\pi a$ ($2m_{\pi}\leq\sqrt{s}\leq m_{\eta^\prime}$), assuming a massless ALP. }
\label{Fig:PhaseInput} 
\end{figure}

Fig.~\ref{Fig:PhaseInput} shows our phase shift input assuming \eqref{Eq:PhaseContinuation}, labelled as the central solution (solid curve). We have checked that different choices for continuation prescriptions, {\it{e.g.}}, following refs.~\cite{DeTroconiz:2001rip,Gonzalez-Solis:2019iod}, result in nearly the same amplitude behavior at low energies, especially in the decay region for the processes studied here.
The behavior of the partial wave in the vicinity of the $K\bar{K}$ threshold, however, is rather ambiguous and raises the question: is the partial wave expected to have a peak or a dip around $K\bar{K}$?
As this question is not easy to answer\footnote{One could attempt to answer this question with a full $\pi\pi$ and $K\bar{K}$ coupled-channel analysis, in the spirit of Ref.~\cite{Albaladejo:2017hhj} for $\eta\to3\pi$. This, however, lies beyond the scope of this work.}, we will quantify uncertainties around and above the inelastic threshold by considering two different scenarios depending on how the partial wave of the system couples to $K\bar{K}$.
On the one hand, the partial wave is expected to have a sharp peak around the position of the $f_{0}(980)$ if it couples strongly to $K\bar{K}$, in which case the phase shift is increased by about $\pi$ while passing through the resonance.
This scenario is realized in the $\pi\pi$ scattering phase shift~\cite{Garcia-Martin:2011iqs}, as well as in the strange scalar form factor of the pion~\cite{Ananthanarayan:2004xy,Oller:2007xd}, and this feature is captured by solution $a$ (dashed curve) in Fig.~\ref{Fig:PhaseInput}. 
On the other hand, a dip in the partial wave is expected if the system couples weakly to strangeness, in which case the phase quickly drops by about $\pi$ with respect to the elastic approximation.
This scenario is realized in the non-strange scalar form factor of the pion~\cite{Ananthanarayan:2004xy,Oller:2007xd}, and this behavior is captured by solution $b$ (dotted curve) in Fig.~\ref{Fig:PhaseInput}.

\begin{figure}
\centering\includegraphics[width=0.95\textwidth]{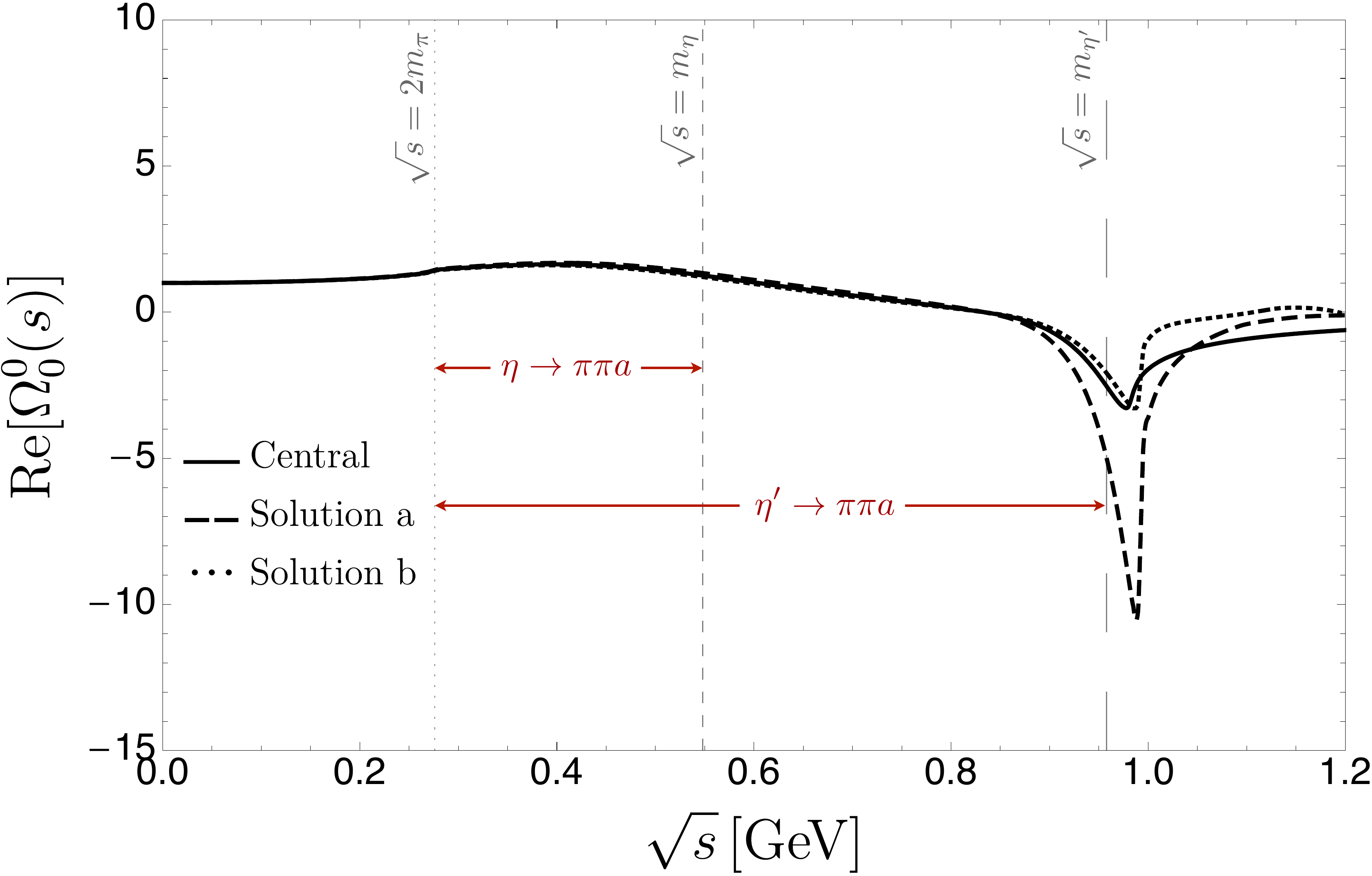}~~~~~\\
\vspace{3em}
\centering\includegraphics[width=0.95\textwidth]{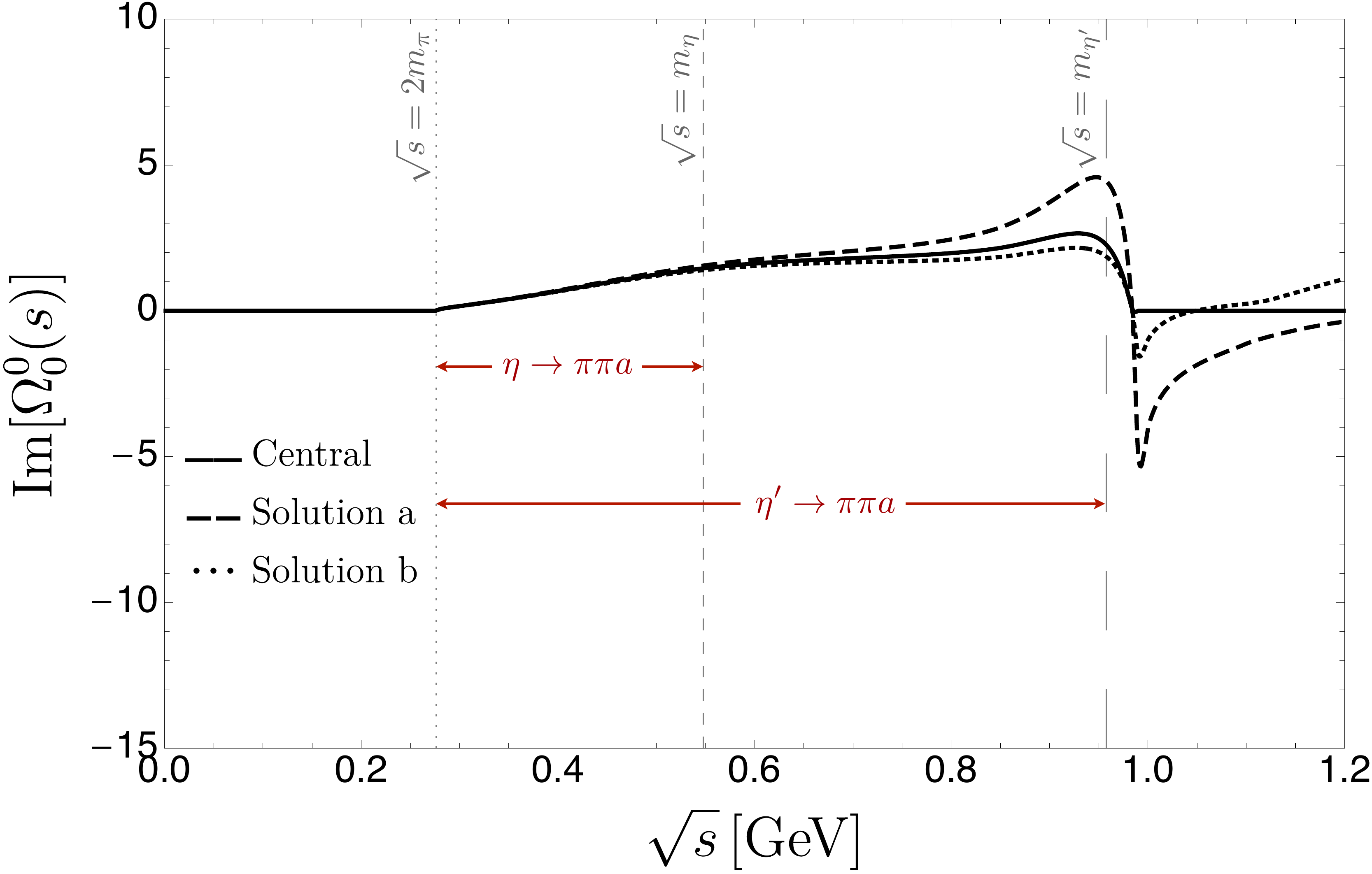}~~~~~
\vspace{2em}
\caption{Real (top) and imaginary (bottom) parts of the $S$-wave Omn\`{e}s functions (\ref{Eq:Omnes}) using the three phase shift examples of Fig.~\ref{Fig:PhaseInput} as inputs. The red arrows indicate the phase space for the decays $\eta^{(\prime)}\to\pi\pi a$ assuming a massless ALP.}
\label{Fig:OmnesSolutions} 
\end{figure}

Solutions $a$ and $b$ for the $\pi\pi$ phase shifts in Fig.\,\ref{Fig:PhaseInput} follow the smoothly interpolating functions devised in Ref.~\cite{Colangelo:2015kha}:
\begin{eqnarray}
\delta_{0}^{0}(s)\big|_{\rm{Solution\,a}}&=&\left(1-f_{\rm{int}}(s_{1},s_{2},s)\right)\delta_{0}^{0}(s)\big|_{\eqref{Eq:PhaseContinuation}}+f_{\rm{int}}(s_{1},s_{2},s)\pi\,,\\
\delta_{0}^{0}(s)\big|_{\rm{Solution\,b}}&=&\left(1-f_{\rm{int}}(s_{1},s_{2},s)\right)\left(\delta_{0}^{0}(s)\big|_{\eqref{Eq:PhaseContinuation}}-f_{\rm{int}}(\tilde{s}_{1},\tilde{s}_{2},s)\pi\right)+f_{\rm{int}}(s_{1},s_{2},s)\pi,~~~~~~~
\end{eqnarray}
where
\begin{eqnarray}
f_{\rm{int}}(s_{1},s_{2},s)=\left\{ \begin{array}{ll}
         0& \mbox{if $s<s_{1}$},\\
         \frac{(s-s_{1})^{2}(3s_{2}-2s-s_{1})}{(s_{2}-s_{1})^{3}}& \mbox{if $s_{1}\le s<s_{2}$},\\
       1& \mbox{if $s\geq s_{2}$}.\end{array} \right.
\end{eqnarray}
In particular, for Fig.\,\ref{Fig:PhaseInput} we assume $s_{1}=68m_{\pi}^{2}$, $s_{2}=105m_{\pi}^{2}$, $\tilde{s}_{1}=4m_{K}^{2}$, and $\tilde{s}_{2}=\tilde{s}_{1}+16m_{\pi}^{2}$, although these values could be varied slightly without any significant consequence.

In Fig.\,\ref{Fig:OmnesSolutions} we show the output for the real and imaginary parts of the $S$-wave Omn\`{e}s function \eqref{Eq:Omnes} using the phase shifts of Fig.\,\ref{Fig:PhaseInput} as inputs.
Note that the differences between the three Omn\`{e}s functions are negligible in the $\eta\to\pi\pi a$ decay region (i.e., for $\sqrt{s}\leq m_\eta$), but become non-negligible near the phase space edge of $\eta^{\prime}\to\pi\pi a$ (i.e., near $\sqrt{s}=m_{\eta^\prime}$).

Finally, for the $\eta^{\prime}\to\eta\pi^{0}a$ channel, we account for the $\eta\pi^{0}$ final state interactions following a treatment analogous to the $\pi\pi$ case described above. Specifically, following (\ref{Eq:GeneralOmnes}), we multiply the leading order amplitude $\mathcal{A}(\eta^{\prime}\to\eta\pi^{0}a)$ given in (\ref{Eq:AmplitudeEtaPrimetoEta}) by the isospin $I=0$ $\eta\pi^{0}$ $S$-wave Omn\`{e}s function $\Omega_{0}^{1}(s)$ (cf.~(\ref{Eq:Omnes})).
To obtain the $\eta\pi^{0}$ phase shift $\delta_{0}^{1}(s)$, and consequently $\Omega_{0}^{1}(s)$, we take the output $S$-wave phase for $\eta\pi^{0}\to\eta\pi^{0}$ scattering obtained in \cite{Gonzalez-Solis:2018xnw}, which uses a unitarized large-$N_{c}$ $\chi$PT treatment to study the decay $\eta^{\prime}\to\eta\pi^{0}\pi^{0}$.

\section{Branching ratios for single ALP channels}\label{sec:pheno}

With our results for the leading order amplitudes and the FSI corrections obtained in Secs.\,\ref{sec:formalism} and \ref{sec:rescattering}, respectively, we can now easily extract the branching ratios for the single ALP decays $\eta^{(\prime)}\to\pi\pi a$ and $\eta^{\prime}\to\eta\pi a$.

For concreteness, we shall adopt two benchmark scenarios for the effective hadronic ALP couplings that have been commonly considered in the literature:
\begin{description}
\item[Gluon-dominance scenario]\hfill\break
This scenario can be loosely thought of as a generalized version of the KSVZ axion \cite{Kim:1979if, Shifman:1979if} for ALPs: the PQ symmetry acts upon some BSM sector carrying SU$(3)$ charges that lives above the EW scale. The SM quarks are uncharged under the PQ symmetry, i.e., 
\begin{equation}
Q_q=0~~\text{for}~~q=u,d,s,c,b,t.
\end{equation}
Once the PQ symmetry is explicitly and spontaneously broken and the heavy states are integrated out, both a PQ-breaking ALP mass term and an ALP-gluon coupling are generated,
\begin{equation}
\mathcal{L}^{Y}_{\rm{ALP}}~\supset~-\frac{1}{2}\MPQ^{2}a^{2}-Q_{G}\frac{\alpha_{s}}{8\pi}\frac{a}{f_{a}}G\tilde{G},
\end{equation}
with
\begin{equation}
Q_G=Q_{G}^{\rm{UV}}\equiv Q.
\end{equation} 
\item[Quark-dominance scenario]\hfill\break
This scenario more closely resembles a generalized version of the DFSZ axion \cite{Dine:1981rt, Zhitnitsky:1980tq} for ALPs, in which all SM quarks carry flavor-universal PQ charges, and there are no additional contributions to the ALP gluon coupling from UV physics (i.e., $Q_{G}^{\rm{UV}}=0$), such that
\begin{subequations}
\begin{alignat}{3}
&Q_q&&\equiv&&~Q~~\text{for}~~q=u,d,s,c,b,t,\\
&Q_G&&=&&~Q_c+Q_b+Q_t=3Q.
\end{alignat}
\end{subequations}
\end{description}

\begin{figure}
\centering\includegraphics[width=0.75\textwidth]{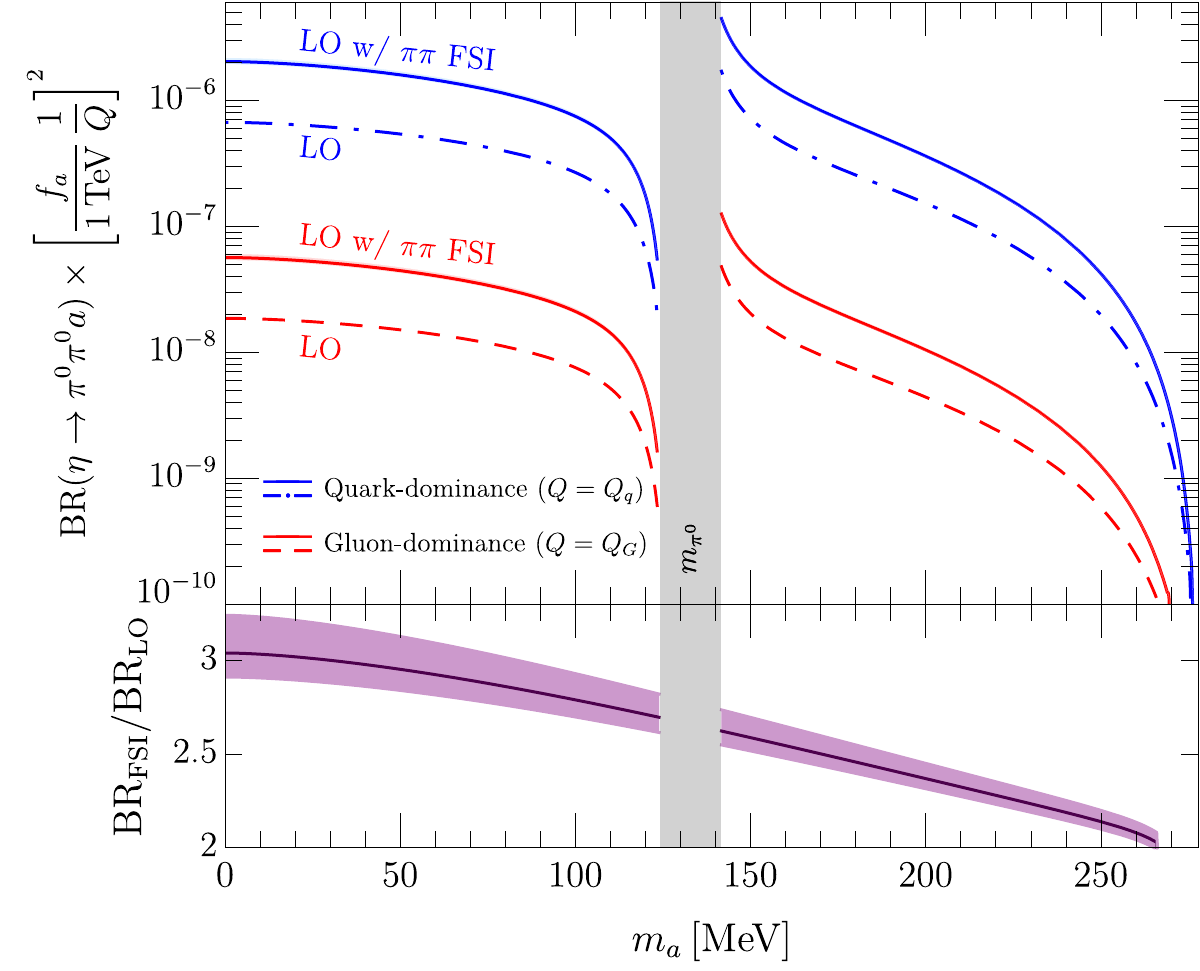}~~~~~\\
\vspace{1em}
\centering\includegraphics[width=0.75\textwidth]{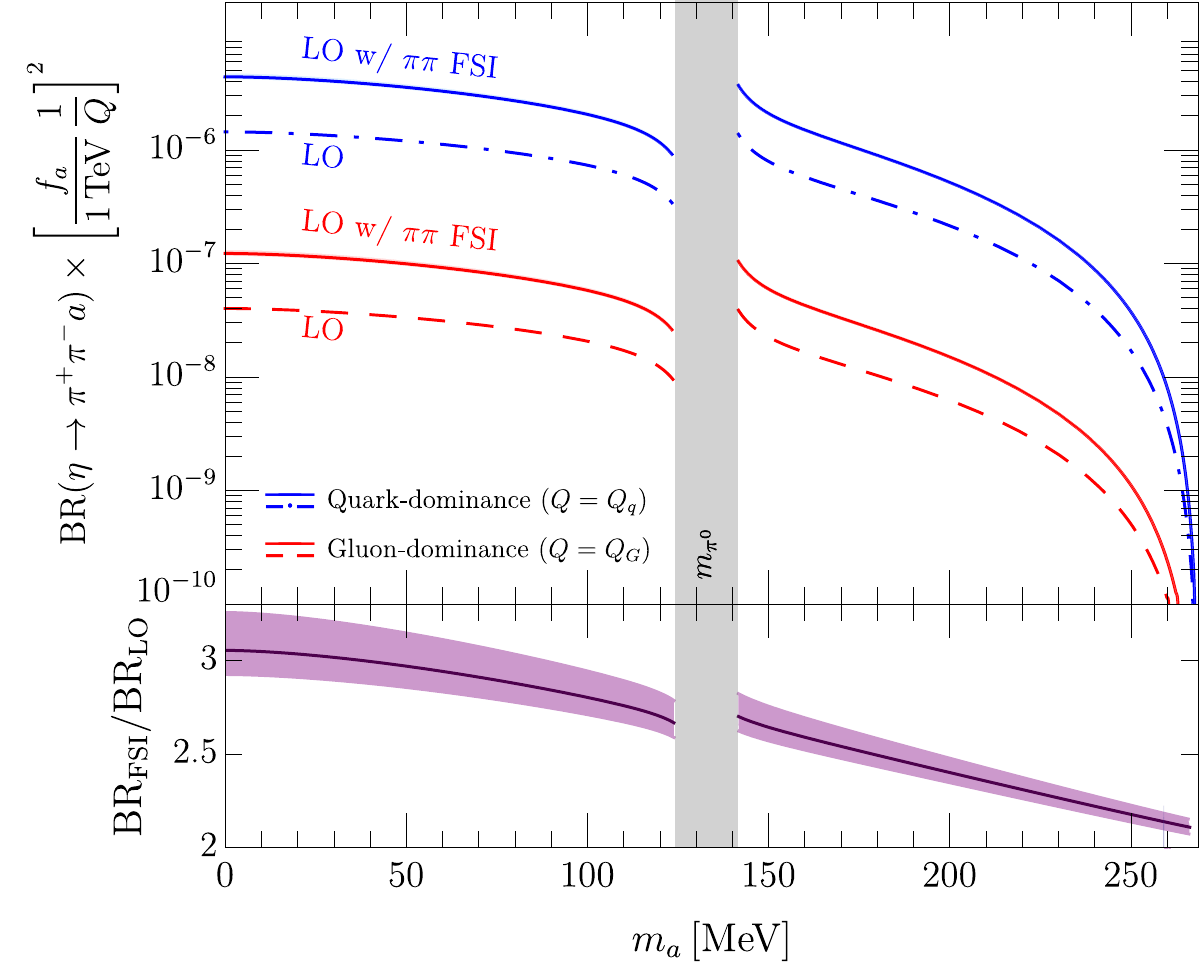}~~~~~
\caption{Branching ratios for $\eta\to\pi^{0}\pi^{0}a$ (top plot) and $\eta\to\pi^{+}\pi^{-}a$ (bottom plot), as a function of the ALP mass $m_a$, for the quark-dominance (solid blue curve) and gluon-dominance (solid red curve) scenarios, including corrections from $\pi\pi$ final state interactions (FSI). For comparison, the corresponding leading order (LO) predictions are indicated by the (dot-)dashed curves.
The bottom panels indicate the overall enhancement of the branching ratios stemming from FSI corrections relative to the LO predictions. The curves' error bands reflect the uncertainties associated with the $\pi\pi$ rescattering phase shift $\delta_{0}^{0}(s)$ (see Subsec.\,\ref{sec:phaseshift}).
Since our small mixing approximations are not valid when $m_{a}\approx m_{\pi^{0}}$ (see Subsec.\,\ref{sec:ALPLagrangian}), this region is masked out in the plots.}
\label{Fig:EtaPredictions} 
\end{figure}
\begin{figure}
\centering\includegraphics[width=0.75\textwidth]{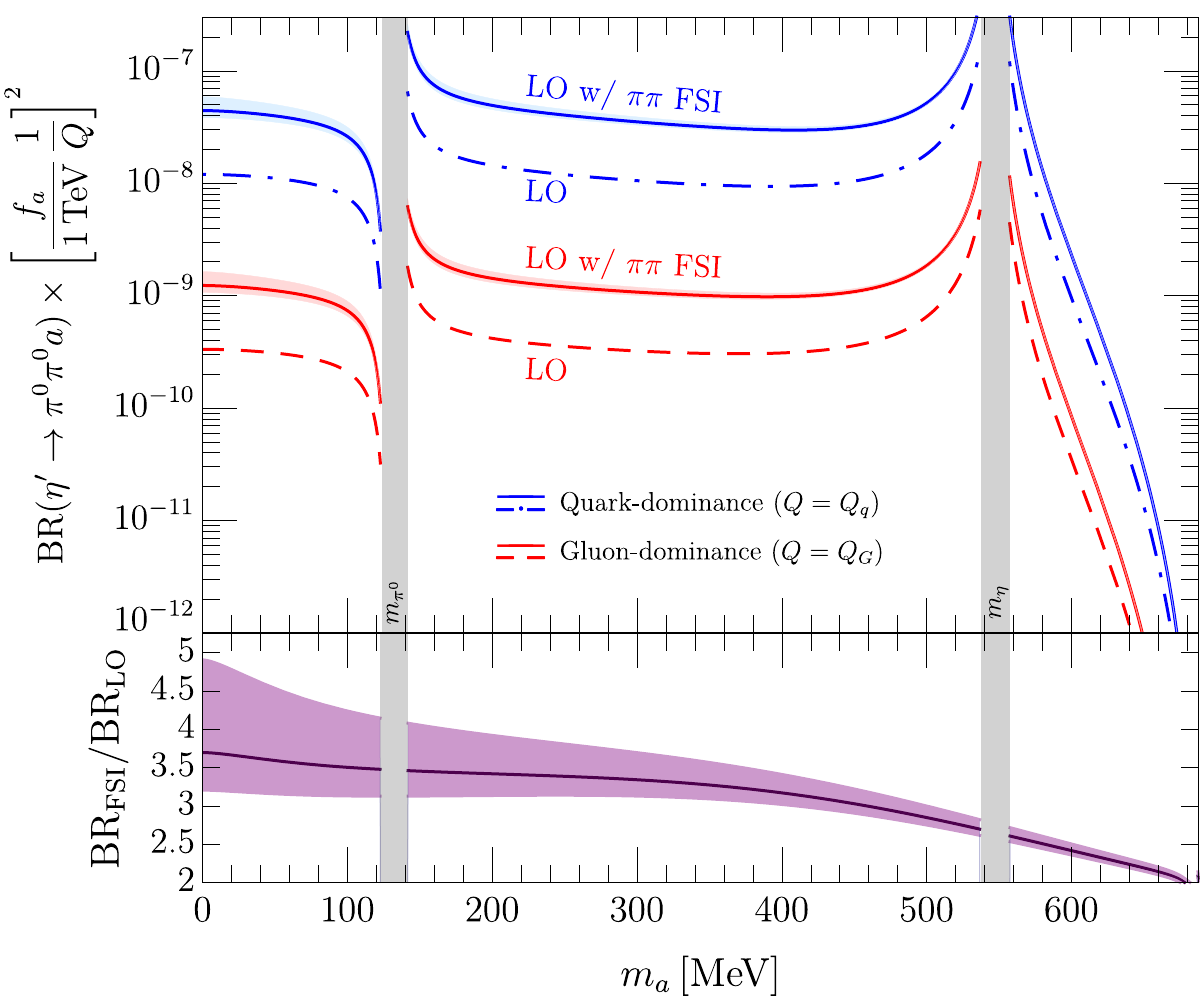}~~~~~\\
\vspace{1em}
\centering\includegraphics[width=0.75\textwidth]{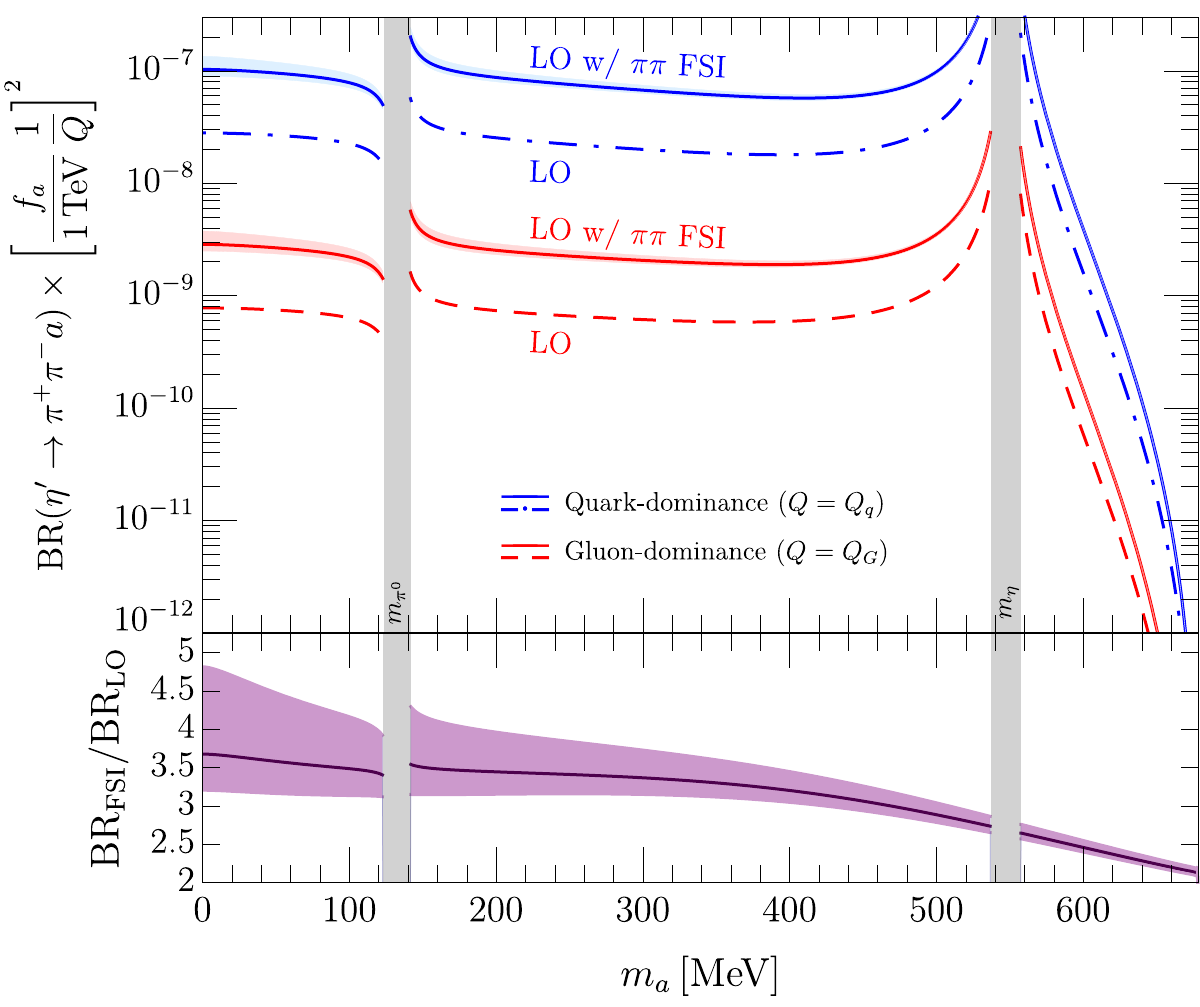}~~~~~
\caption{Branching ratios for $\eta^{\prime}\to\pi^{0}\pi^{0}a$ (top plot) and $\eta^{\prime}\to\pi^{+}\pi^{-}a$ (bottom plot), as a function of the ALP mass $m_a$, for the quark-dominance (solid blue curve) and gluon-dominance (solid red curve) scenarios, including corrections from $\pi\pi$ final state interactions (FSI). For comparison, the corresponding leading order (LO) predictions are indicated by the (dot-)dashed curves.
The bottom panels indicate the overall enhancement of the branching ratios stemming from FSI corrections relative to the LO predictions. The curves' error bands reflect the uncertainties associated with the $\pi\pi$ rescattering phase shift $\delta_{0}^{0}(s)$ (see Subsec.\,\ref{sec:phaseshift}).
Since our small mixing approximations are not valid when $m_{a}\approx m_{\pi^{0}}$ and $m_{a}\approx m_{\eta}$ (see Subsec.\,\ref{sec:ALPLagrangian}), these regions are masked out in the plots.}
\label{Fig:EtapPredictions} 
\end{figure}
\begin{figure}
\centering\includegraphics[width=0.70\textwidth]{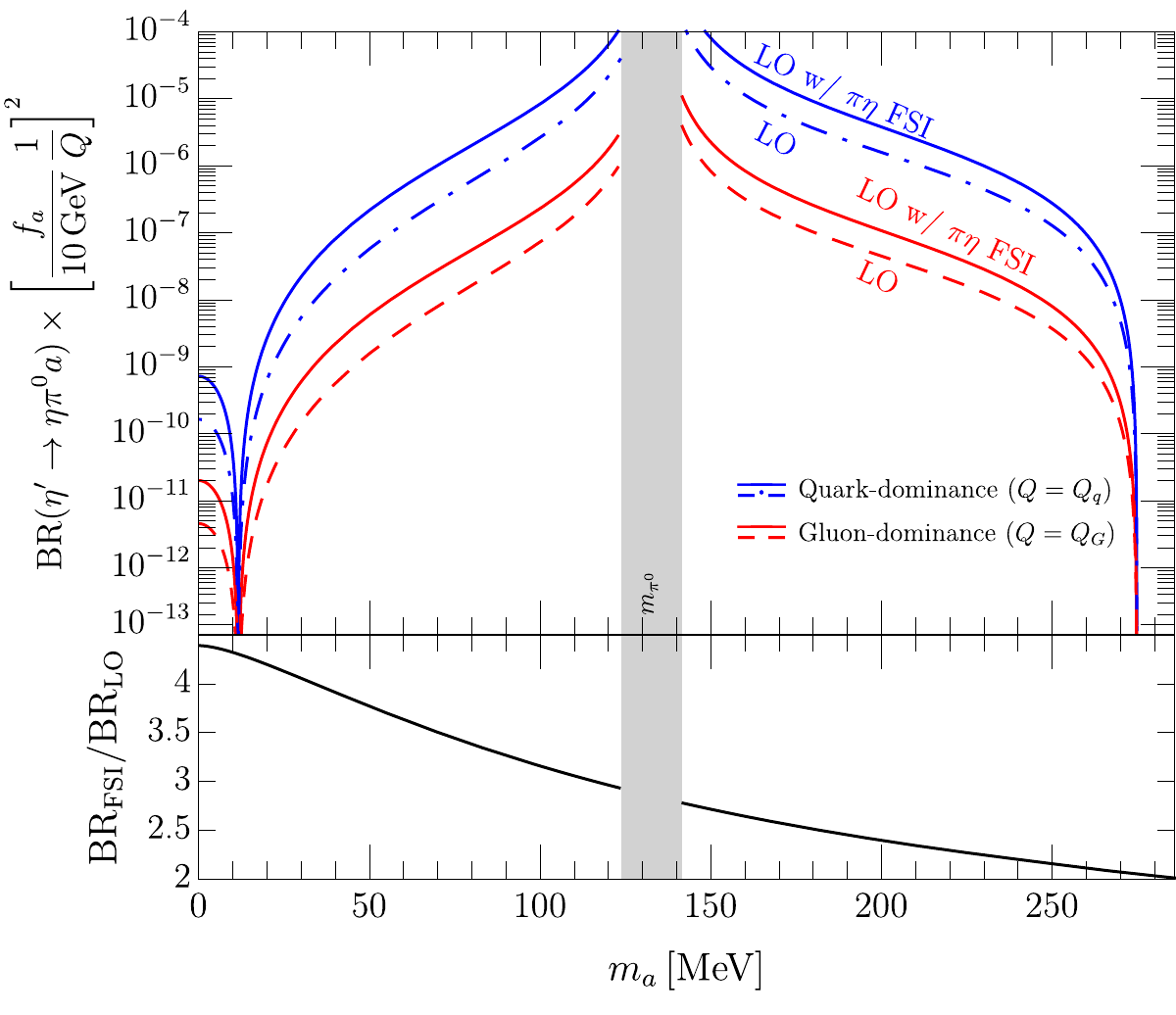}~~~~~
\caption{Branching ratio for $\eta^{\prime}\to\eta\pi^{0}a$, as a function of the ALP mass $m_a$, for the quark-dominance (solid blue curve) and gluon-dominance (solid red curve) scenarios, including corrections from $\eta\pi^{0}$ final state interactions (FSI). For comparison, the corresponding leading order (LO) predictions are indicated by the dashed curves.
The bottom panel indicates the overall enhancement of the branching ratio stemming from FSI corrections relative to the LO prediction. Since our small mixing approximations are not valid when $m_{a}\approx m_{\pi^{0}}$, this region is masked out in the plot.}
\label{Fig:EtaPtoEtaPredictions} 
\end{figure}

In Figs.~\ref{Fig:EtaPredictions}, \ref{Fig:EtapPredictions}, and \ref{Fig:EtaPtoEtaPredictions}, we show the branching ratios for the decays $\eta^{(\prime)}\to\pi^{0}\pi^{0}a$, $\eta^{(\prime)}\to\pi^{+}\pi^{-}a$, and $\eta^{\prime}\to\eta\pi^{0}a$ as a function of the ALP mass $m_{a}$ for both the quark-dominance and gluon-dominance benchmark scenarios.
The corrections due to pion-pion final state interactions are shown at the bottom panel of these figures: the curves indicate the overall enhancement of the branching ratio relative to the leading order estimate as a function of the ALP mass $m_{a}$. The error bands indicate the uncertainties associated with the $\pi\pi$ rescattering phase shift $\delta_{0}^{0}(s)$ (cf.~Figs.~\ref{Fig:PhaseInput} and \ref{Fig:OmnesSolutions}).
As seen, the strong rescattering effects enhance these axionic decay rates by factors ranging from $\sim 2-3$ over the kinematically allowed ALP mass range for $\eta$ decays, and by factors of $\sim 2-4$ for $\eta^{\prime}$ decays. This clearly supports our initial argument that ALP studies based on leading order treatments, commonly seen in the literature, fall short of correctly capturing the phenomenology of hadronic ALPs, and more rigorous treatments including strong interaction effects are sorely needed.

\section{Multi-ALP final states}\label{sec:MultiALP}

In addition to the single ALP processes just considered, it is also worth exploring $\eta^{(\prime)}$ decay channels with multiple ALPs in the final state. Generically, the rate for emission of $n_a$ ALPs is suppressed by a factor of $(1/f_a)^{2n_a}$, where $f_a$ is the ALP decay constant. Hence, for models with high scale PQ symmetry breaking, and hence with high $f_a$, multi-ALP channels are hopelessly outside the reach of rare $\eta^{(\prime)}$ decay searches. However, in models with $f_{a}$ below the EW scale, the branching ratios for multi-ALP channels might be experimentally accessible, and in fact even competitive with single ALP channels due to much more suppressed backgrounds. While viable ALP models with $f_{a}<v_{\text{\tiny EW}}$ are difficult to model-build, examples exist in the literature (\cite{Alves:2017avw,Alves:2020xhf,Liu:2021wap}).

It is straightforward to obtain the leading order amplitudes for double- and triple-ALP decay channels, such as $\eta^{(\prime)}\to\pi^{0}aa$, $\eta^{\prime}\to\eta aa$, and $\eta^{(\prime)}\to aaa$.
Using (\ref{Eq:Mixing}) and expanding the Lagrangian (\ref{Eq:Lag}) in the physical ALP field to the appropriate order, we find:
\begin{eqnarray}
\mathcal{A}(\eta\to\pi aa)&=&\frac{m_\pi^2}{f_\pi^2}\,C_{\eta}\,\frac{(m_uA_u^2-m_dA_d^2)}{m_u+m_d},\label{Eq:DoublyALPeta}\\
\mathcal{A}(\eta^\prime\to\pi aa)&=&\frac{m_\pi^2}{f_\pi^2}\,C_{\eta^\prime}\,\frac{(m_uA_u^2-m_dA_d^2)}{m_u+m_d},\label{Eq:DoublyALPetaP}\\
\mathcal{A}(\eta^{\prime}\to\eta aa)&=&\frac{m_\pi^2}{f_\pi^2}\,C_{\eta^\prime}C_{\eta}\,\frac{(m_uA_u^2+m_dA_d^2-4m_sA_s^2)}{m_u+m_d},\label{Eq:etaPetaAA}\\
\mathcal{A}(\eta\to aaa)&=&\frac{m_\pi^2}{f_\pi^2}\,\;\frac{C_{\eta}(m_uA_u^3+m_dA_d^3)-4C_{\eta^\prime}m_sA_s^3}{m_u+m_d},\label{Eq:TriplyALPeta}\\
\mathcal{A}(\eta^\prime\to aaa)&=&\frac{m_\pi^2}{f_\pi^2}\;\frac{C_{\eta^\prime}(m_uA_u^3+m_dA_d^3)+4C_{\eta}m_sA_s^3}{m_u+m_d}.\label{Eq:TriplyALPetaP}
\end{eqnarray}
In \eqref{Eq:DoublyALPeta}-\eqref{Eq:TriplyALPetaP} above we have omitted the small contributions proportional to $\theta_{\pi\eta^{(\prime)}}$, and the variables $C_{\eta}$, $C_{\eta^{\prime}}$, and $A_{q=u,d,s}$ have been defined in \eqref{Ceta}–\eqref{As}. Note that because the ALP-meson mixing angles, as well as the variables $A_{q=u,d,s}$, scale as $1/f_a$, the parametric dependence of the $\eta^{(\prime)}$ decay amplitudes on the ALP decay constant scales as $(1/f_a)^{n_a}$, where $n_a$ is the number of ALPs in the final state.

\begin{figure}
\centering\includegraphics[width=0.7\textwidth]{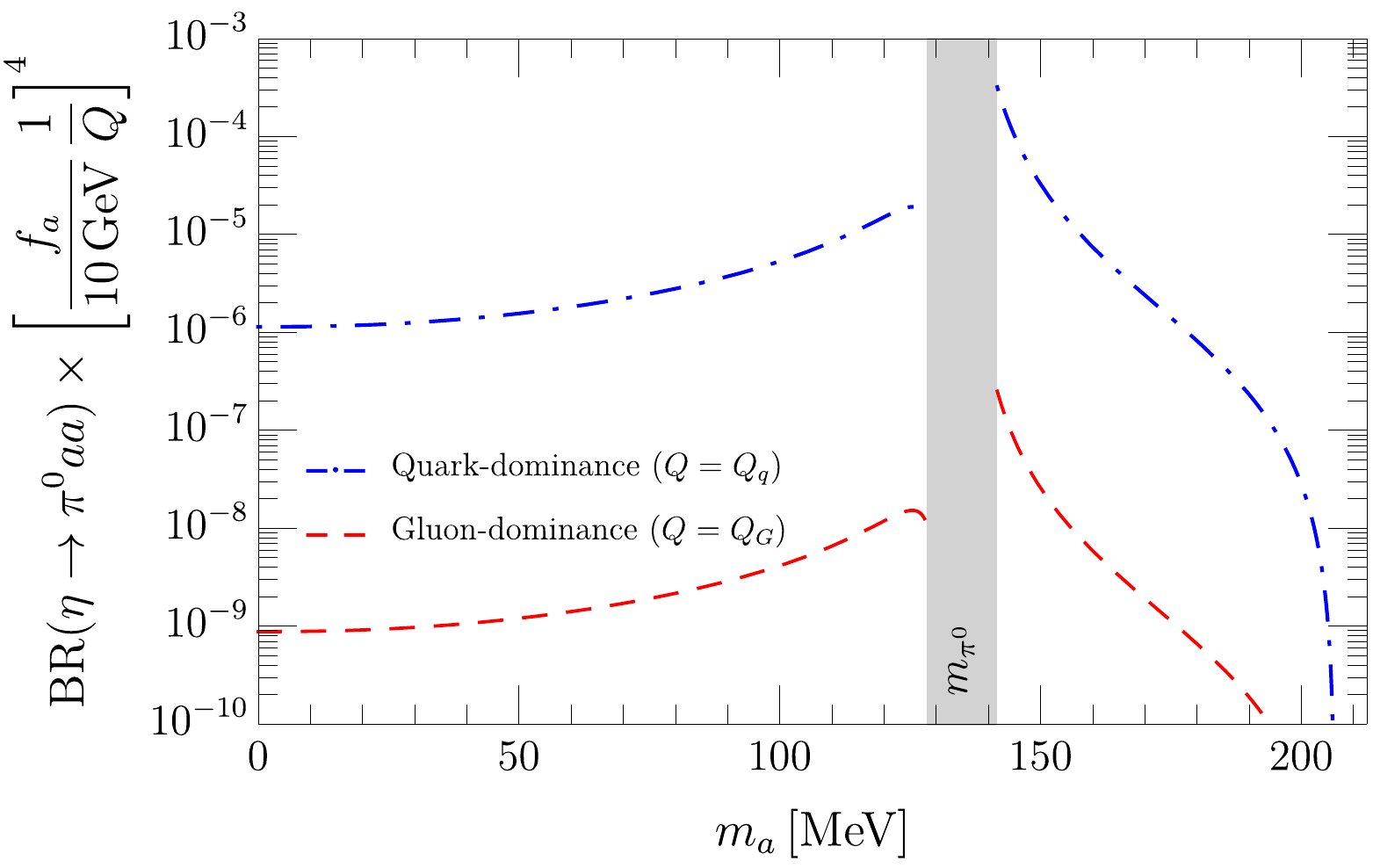}~~~~~\\
\vspace{1em}
\centering\includegraphics[width=0.7\textwidth]{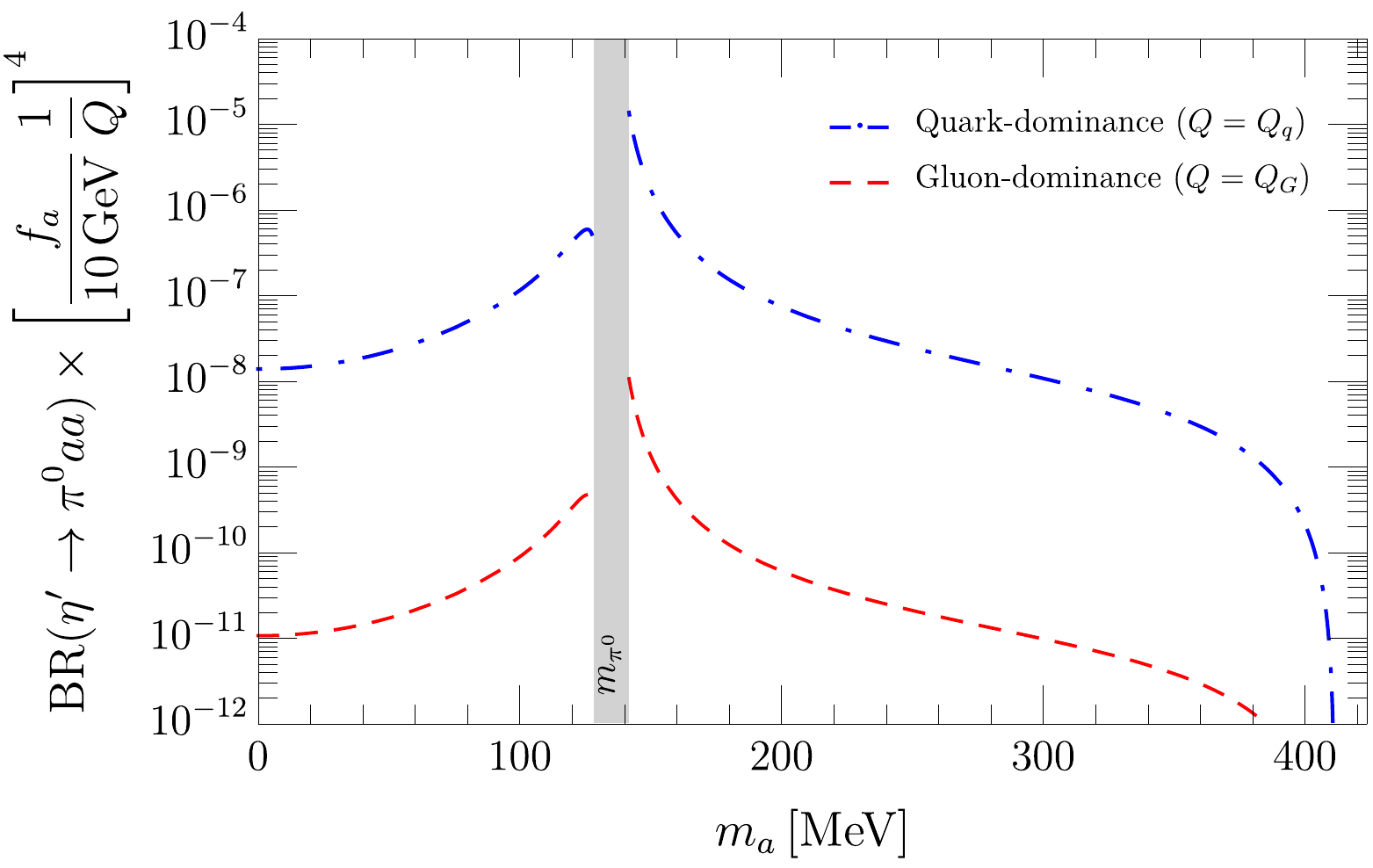}~~~~~\\
\vspace{1em}
\centering\includegraphics[width=0.7\textwidth]{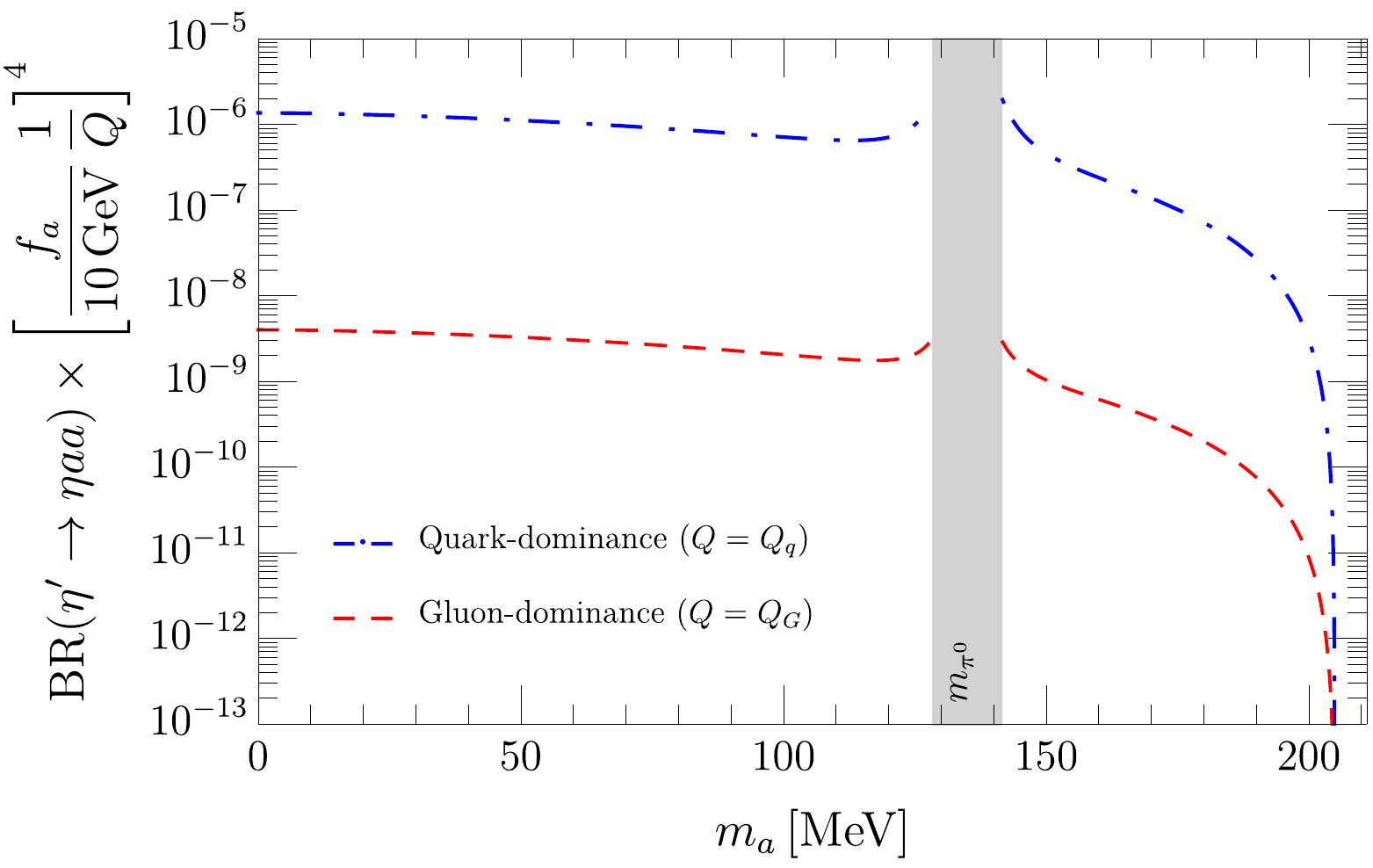}~~~~~
\caption{
Leading order branching ratios for the the double-ALP channels $\eta\to\pi^{0}aa$ (top plot), $\eta^{\prime}\to\pi^{0}aa$ (middle plot), and $\eta^{\prime}\to\eta aa$ (bottom plot), as a function of the ALP mass $m_a$, for the quark-dominance (dot-dashed blue curve) and gluon-dominance (dashed red curve) scenarios. Since our small mixing approximations are not valid when $m_{a}\approx m_{\pi^{0}}$ (see Subsec.\,\ref{sec:ALPLagrangian}), this region is masked out in the plots.}
\label{Fig:TwoALPpredictions} 
\end{figure}

\begin{figure}
\centering\includegraphics[width=0.7\textwidth]{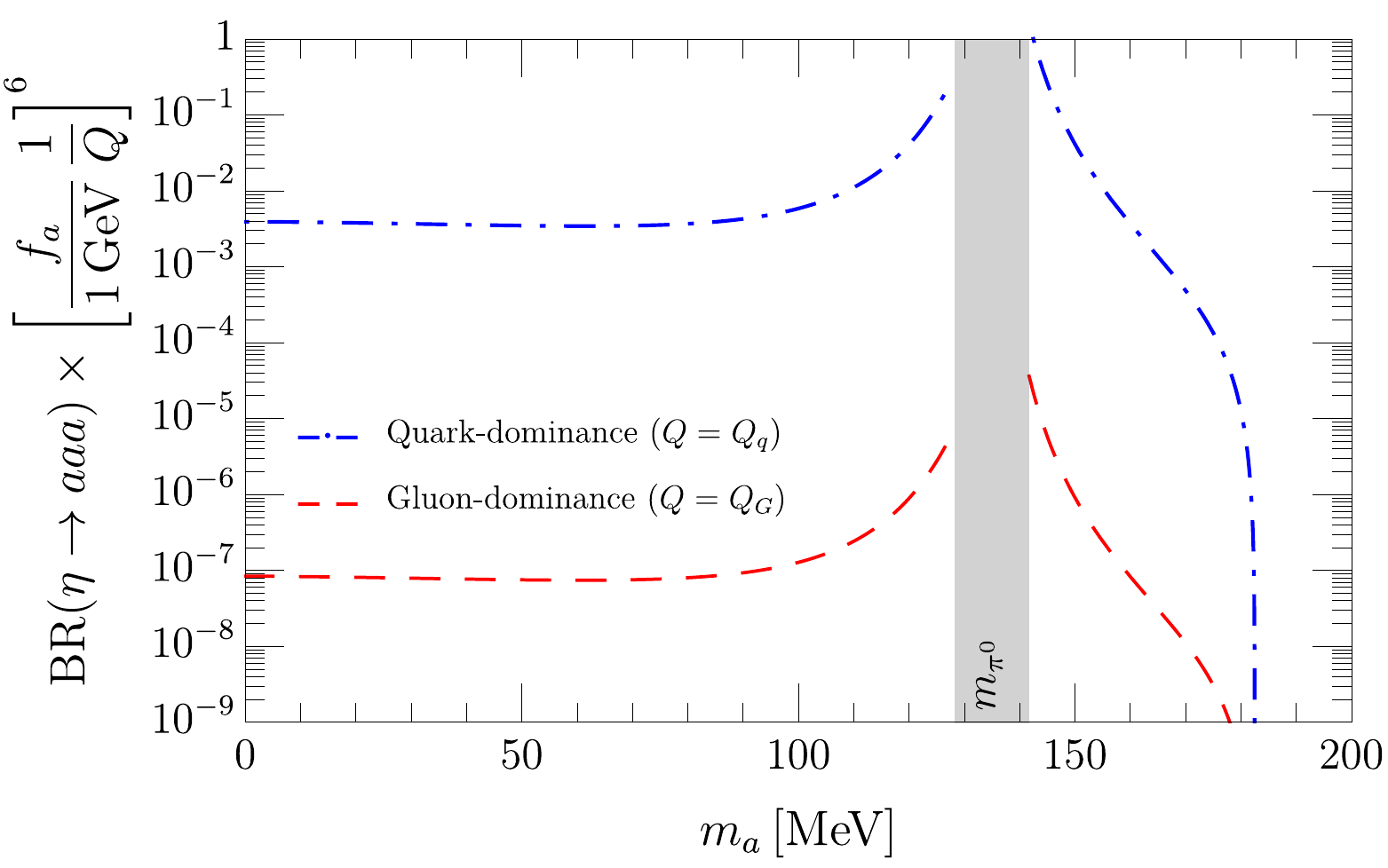}~~~~~\\
\vspace{1em}
\centering\includegraphics[width=0.7\textwidth]{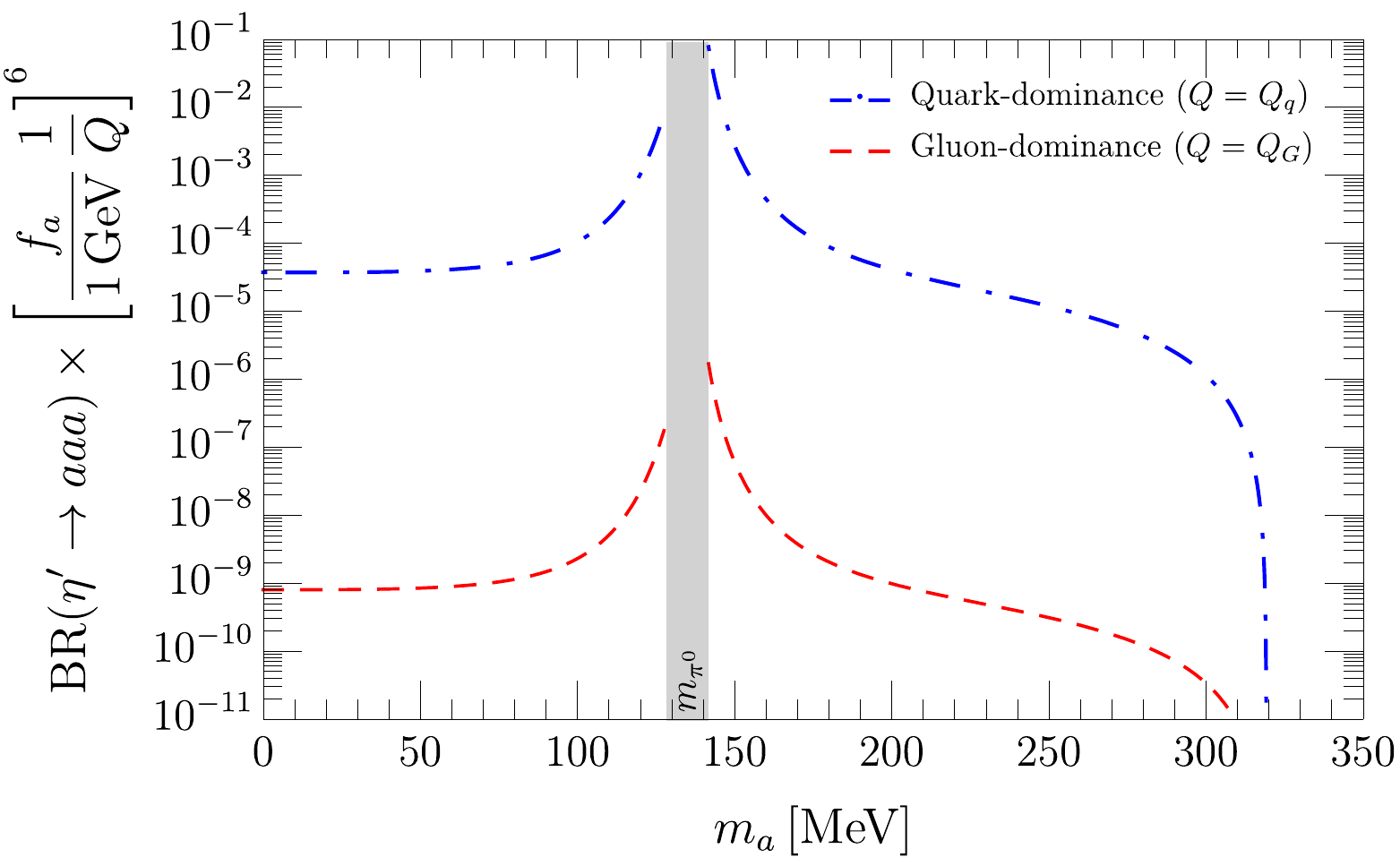}~~~~~
\caption{Leading order branching ratios for the the triple-ALP channels $\eta\to aaa$ (top plot) and $\eta^{\prime}\to aaa$ (bottom plot), as a function of the ALP mass $m_a$, for the quark-dominance (dot-dashed blue curve) and gluon-dominance (dashed red curve) scenarios. Since our small mixing approximations are not valid when $m_{a}\approx m_{\pi^{0}}$ (see Subsec.\,\ref{sec:ALPLagrangian}), this region is masked out in the plots.}
\label{Fig:ThreeALPpredictions} 
\end{figure}

In Fig.~\ref{Fig:TwoALPpredictions}, we show the branching ratios for the double-ALP decay channels $\eta^{(\prime)}\to\pi^{0}aa$ and $\eta^{\prime}\to\eta aa$, as a function of the ALP mass $m_{a}$, for both the quark-dominance and gluon-dominance benchmark scenarios. Note that in these plots we have normalized the branching ratios to a low decay constant of $f_{a}=10$ GeV, in order to display decay rates that are potentially  accessible to experiments in current and future $\eta/\eta^\prime$ factories.
We show analogous plots in Fig.~\ref{Fig:ThreeALPpredictions} for the triple-ALP decay channels $\eta^{(\prime)}\to aaa$, this time normalized to an even lower decay constant of $f_{a}=1$ GeV. 

\begin{figure}
\centering\includegraphics[width=0.95\textwidth]{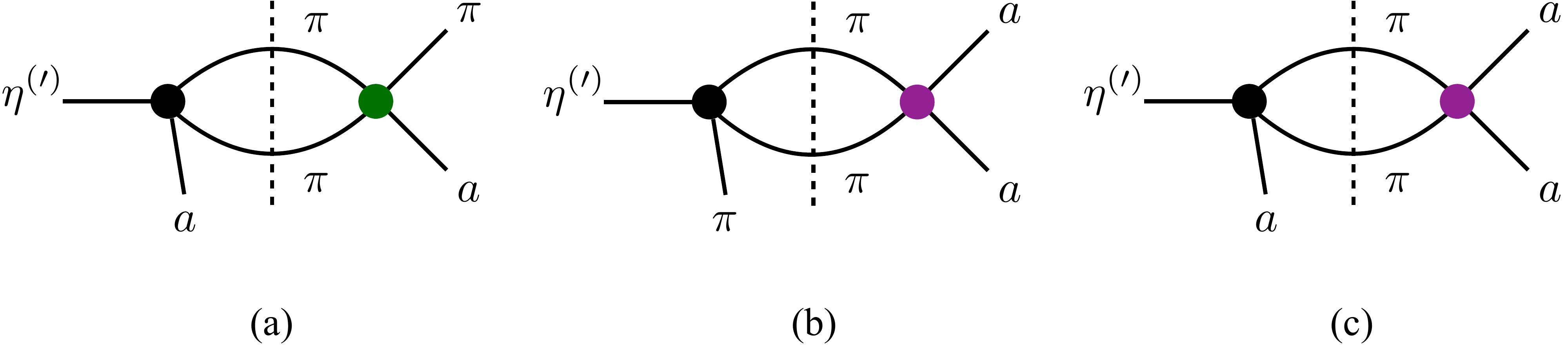}
\caption{Schematic representation of the $\pi\pi$ rescattering contributions to the double-ALP channel $\eta^{(\prime)}\to\pi^{0}aa$ (diagrams (a) and (b)), and the triple-ALP channel $\eta^{(\prime)}\to aaa$ (diagram (c)).}
\label{Fig:MultiALP} 
\end{figure}

As with the single ALP processes, we expect that these leading order predictions will receive sizable unitarity corrections from $\pi\pi$ intermediate states. Fig.\,\ref{Fig:MultiALP} illustrates a few processes that can enhance double- and triple-ALP decay channels. In particular, diagrams (a) and (b) illustrate, respectively, the contributions of $\pi\pi\to\pi a$ and $\pi\pi\to aa$ rescattering to the double-ALP channel $\eta^{(\prime)}\to\pi^{0}aa$. Even neglecting complications from $3\pi$ rescattering, our results from the analysis in Sec.\,\ref{sec:rescattering} indicate that $\pi\pi$ rescattering in the $s$-channel alone should provide a rate enhancement over the leading order expectation by at least a factor of 2. Similarly, diagram (c) illustrates the contribution of $\pi\pi\to aa$ rescattering to the triple-ALP channel $\eta^{(\prime)}\to aaa$, from which we expect an enhancement of the decay rates by a factor of $\sim2.5-3$. A complete analysis of FSI corrections to multi-ALP channels in $\eta$, $\eta^{\prime}$, and other meson decays is deferred to a future publication.

\section{Conclusions and Outlook}\label{sec:summary}

In this work, we have performed a theoretical analysis of axio-hadronic decays of the $\eta$ and $\eta^{\prime}$ mesons, paying particular attention to single ALP emission in $\eta^{(\prime)}\to\pi\pi a$ decays, where $\pi\pi=\pi^{0}\pi^{0},\pi^{+}\pi^{-}$. 
Using the framework of U(3) chiral perturbation theory, we have calculated the leading order amplitudes for these processes, and using dispersion relations, we have accounted for the strong $\pi\pi$ final state interactions in a model-independent way, a significant improvement with respect to existing literature.
The results of our calculations of the branching ratios as a function of the ALP mass are given in Figs.~\ref{Fig:EtaPredictions} and \ref{Fig:EtapPredictions}.
We have shown that $\pi\pi$ rescattering is important in order not to underestimate these branching ratios.
In particular, this effect enhances the decay rates by factors that can be as large as 3 for axio-hadronic $\eta$ decays and as large as 4 for axio-hadronic $\eta^{\prime}$ decays.
The inclusion of final state rescattering effects is the main result of this work; further improvements to include additional next-to-leading order contributions could refine the predictions presented here. We leave this investigation for future work.

We have also provided, for the first time, leading order predictions for double- and triple-ALP emission in $\eta$ and $\eta^\prime$ decays, including $\eta^{(\prime)}\to\pi^{0}aa$, $\eta^{\prime}\to\eta aa$ and $\eta^{(\prime)}\to aaa$ (see Figs.~\ref{Fig:TwoALPpredictions} and \ref{Fig:ThreeALPpredictions}). These channels can be explored in experimental searches to probe low scale PQ symmetry breaking scenarios, for which the ALP decay constant lies below the electroweak scale.

With upcoming $\eta/\eta^\prime$ factories in the horizon, dedicated searches for hadronic ALPs would open a new window into the exploration of strong CP solutions and low scale dark sectors. The JLab Eta Factory~\cite{JEF,Mack:2014gma}, the REDTOP experiment~\cite{REDTOP:2022slw}, and the super $\tau$-charm facility~\cite{Achasov:2023gey} will determine rare $\eta$ and $\eta^{\prime}$ decays with precision several orders of magnitude higher than present measurements, and the HADES experiment at GSI plans to perform a dedicated search for $\eta\to\pi^{+}\pi^{-}(a\to e^{+}e^{-})$~\cite{HADESTrento2023,HADESprivatecorrespondence}.
Also, dedicated searches at BESIII, KLOE, and the CMS B-parking and data scouting datasets \cite{CMS:2023thf,CMSprivatecorrespondence} could probe a variety of unexplored $\eta$ and $\eta^{\prime}$ BSM decay channels.

Our results provide a first step into improving the accuracy of predictions for axio-hadronic $\eta^{(\prime)}$ decay rates by including the leading effects of non-perturbative strong dynamics through dispersion relations. In order to fully map out the space of experimental signatures, our results should be combined with further assumptions about the ALP couplings to leptons and gauge bosons to obtain their decay modes and lifetimes (see also \cite{REDTOP:2022slw}). This is particularly relevant for models of low scale ALP decay constants $f_a\lesssim\mathcal{O}(10-100)$TeV, for which the ALP can decay into visible final states---such as $\gamma\gamma$, $\ell^{+}\ell^{-}$, $\pi\pi\pi$---with reasonably short lifetimes, yielding prompt or displaced vertices in the detector. Comprehensive phenomenological studies of all possible decay signatures to inform experimental searches are deferred to future work.



\acknowledgments
This work was supported by the DOE Office of Science High Energy Physics under contract number DE-AC52-06NA25396, and by the Laboratory Directed Research and Development program of Los Alamos National Laboratory under project number 20210944PRD2. Los Alamos National Laboratory is operated by Triad National Security, LLC, for the National Nuclear Security Administration of the U.S. Department of Energy under contract number 89233218CNA000001. DSMA acknowledges the Aspen Center for Physics (ACP) where this work was partially developed---the ACP is supported by the National Science Foundation under grant PHY-1607611. SGS is a Serra H\'{u}nter Fellow.



\bibliographystyle{JHEP} 
\bibliography{bibliography}

\end{document}